\documentclass[journal]{IEEEtran}

\usepackage[numbers,sort&compress]{natbib}
\usepackage{url}
\usepackage{bm}
\usepackage{amsfonts}
\usepackage{amsmath}
\usepackage{amssymb}
\usepackage{amsthm}
\usepackage{color}
\usepackage{enumerate}
\usepackage{makecell}
\usepackage{mathrsfs}

\usepackage{array,color}%
\usepackage{tabularx}%
\usepackage{multirow}%
\usepackage{booktabs}%
\usepackage{makecell}  

\usepackage{graphicx}  
\usepackage{subfig}
\usepackage{epstopdf}
\usepackage{graphics}
\usepackage{epsfig}
\usepackage{psfrag}
\usepackage{overpic}

\usepackage{multicol}

 \usepackage{stfloats}
 \usepackage{float}   %

 \usepackage{indentfirst} %
\usepackage{gensymb}
\usepackage{xcolor}%
\usepackage{pifont}

\usepackage{color, colortbl}

\begin{document}
\title{Redefinition of Digital Twin and its Situation Awareness Framework Designing Towards Fourth Paradigm for Energy Internet of Things}
\author{Xing~He,~\IEEEmembership{Senior Member,~IEEE}, Yuezhong~Tang, Shuyan~Ma, Qian~Ai, Fei~Tao, Robert~Qiu,~\IEEEmembership{Fellow,~IEEE},
\thanks{E-mail address of the corresponding author Xing HE: hexing\_hx@126.com. This work was partly supported by National Key R\&D Program of China (2021YFB2401204), National Natural Science Foundation of China (52277111), and Science and Technology Commission of Shanghai Municipality (21DZ1208300).}
}
\maketitle



\begin{abstract}
Traditional knowledge-based situation awareness (SA) modes struggle to adapt to the escalating complexity of today's Energy Internet of Things (EIoT), necessitating a pivotal paradigm shift.
In response, this work introduces a pioneering data-driven SA framework, termed digital twin-based situation awareness (DT-SA), aiming to bridge existing gaps between data and demands, and further to enhance SA capabilities within the complex EIoT landscape.
First, we redefine the concept of digital twin (DT) within the EIoT context, aligning it with data-intensive scientific discovery paradigm (the Fourth Paradigm) so as to waken EIoT's sleeping data; this contextual redefinition lays the cornerstone of our DT-SA framework for EIoT.
Then, the framework is  comprehensively explored through its four fundamental steps: digitalization, simulation, informatization, and intellectualization.
These steps initiate a virtual ecosystem conducive to a continuously self-adaptive, self-learning, and self-evolving big model (BM), further contributing to the evolution and effectiveness of DT-SA in engineering.
Our framework is characterized by the incorporation of system theory and Fourth Paradigm as guiding ideologies, DT as data engine, and BM as intelligence engine. This unique combination forms the backbone of our approach.
This work extends beyond engineering, stepping into the domain of data science---DT-SA not only enhances management practices for EIoT users/operators, but also propels advancements in pattern analysis and machine intelligence (PAMI) within the intricate fabric of  a complex system.
Numerous real-world cases validate our DT-SA framework.
\end{abstract}


\IEEEpeerreviewmaketitle
\section{Introduction}
\label{Sec:Introd}
\begin{figure*}[htbp]
\centerline{
\includegraphics[width=.88\textwidth]{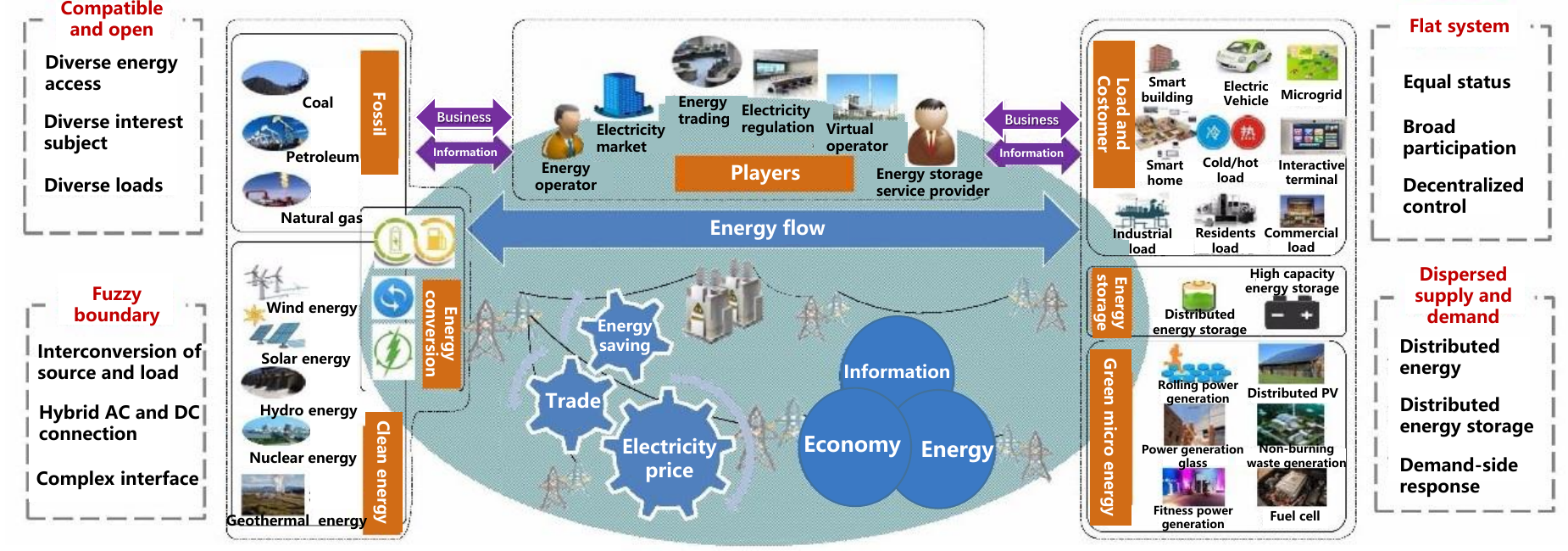}
}
\caption{Development trend of modern EIoT~\cite{guo2019evolution}}
\label{fig:TrendofSG}
\end{figure*}

\IEEEPARstart Energy internet of things (EIoT) emerges as a potential solution to energy crisis and environmental problems.
EIoT encompasses heterogeneous distributed energy resources units (DERs)~\cite{zhou2018distributed}, including uncertain distributed generation units (e.g., photovoltaic, wind turbines), flexible demand-side units (e.g., air conditioning, thermal energy storage), and energy prosumer (e.g., electric vehicles, energy storage).
\textbf{EIoT is a complex system in essence}~\cite{heng2008conflict}, and its complexity is ever increasing in various aspect~\cite{he2020invisible}:
a) connecting nearby utilities improves security and efficiency, resulting in a vast interconnected system;
b) continuous injection of cell units, DERs as typical ones, that are large in number, small in size, distributed in deployment, diverse in behaviors, smart in response, and uncertain in control~\cite{dagoumas2019review};
and c) multiple physical disciplines (electrical, magnetomechanics, etc.) are closely coupled~\cite{fu2017uncertainty}.
All these driving forces lead to an open, flat, and non-linear  complex EIoT as illustrated in Figure~\ref{fig:TrendofSG}~\cite{guo2019evolution}.

For such a complex EIoT, situation awareness (SA) holds critical significance for its routine operation.
Reference~\cite{endsley2000situation} defines SA as the {perception} of environmental elements, the {comprehension} of their meaning, and the {projection} of their status in the near future.
In the realm of EIoT, SA plays a pivotal role in supporting decision-making for both the individual behaviors of DER as well as the collective actions of DER aggregation.
Moreover, SA contributes to the stable operation of EIoT---inadequate SA was identified as a root cause for the world's biggest blackout in history, the US--Canada Blackout~\cite{us2004final}.
This underscores the crucial role of SA in preventing and mitigating potentially catastrophic events in the daily operation of EIoT.

\subsection{Motivation for DT-based Situation Awareness}
\label{sec:motiv}
Traditional SA modes are in urgent need of change, since their struggle to keep pace with EIoT's escalating complexity.
This complexity frequently gives rise to emergence~\cite{holland2000emergence} and chaos~\cite{260940} phenomena \textbf{that go beyond the scope of reductionism\footnote{ Reductionism is any of several related philosophical ideas regarding the associations between phenomena which can be described in terms of other simpler or more fundamental phenomena. It is also described as an intellectual and philosophical position that interprets a complex system as the sum of its parts~\cite{wiki2023Red}.}}, as discussed in Section~\ref{Sec: motivationofAgg}.

Taking a step back, even if reductionism may sometimes effectively model certain EIoT scenarios, the complexity of EIoT modeling often requires addressing  comprehensive and flexible scenarios, taking into account multi-stakeholder dynamics, multi-objective requirements, diverse paths, and high levels of uncertainty.
In this regard, \textbf{both knowledge-based modes and solely data-driven modes exhibit limitations}.
Knowledge-based algorithms are vulnerable to insufficient model accuracy~\cite{shahsavari2019situational} or even the absence of domain-specific model~\cite{magazzino2020relationship}.
Yet, despite a solely data-driven model (e.g., neural network) appearing successfully in building a ``black-box'' for specific purpose (e.g., anomaly detection~\cite{ahmed2021anomaly, li2021fast}), it still struggles to meet SA requirements in EIoT, mainly due to two factors: a) inadequate samples for model training, especially those with ``abnormal operating conditions'' labels, and b) the inherent lack of interpretability for ``black-box'' models.

\subsection{Contribution and Organization of this Work}
Thanks to data-intensive scientific discovery paradigm (i.e., the Fourth Paradigm~\cite{hey2009fourth}), this paper, by integrating the ongoing digitalization progress of EIoT and its overall digital architecture, introduces an SA framework based on our redefined digital twin (DT), termed DT-SA.
DT-SA is designed to systematically explore various potential evolutionary paths of EIoT in the virtual space, with a coverage of the simulation, analysis, decision-making, and verification procedures.
The ever-increasing growth of data in EIoT, coupled with emerging {data technologies/disciplines},  sets the stage for our DT-SA.

The exploration of DT-SA begins by redefining the concept of DT.
Although DT's origins lie in fields like aviation and workshops, \textbf{the DT tailored for EIoT demonstrates distinctive characteristics that deserve a redefinition}.
In contrast to domains heavily reliant on computer vision and precise control~\cite{leng2023manuchain}, EIoT-DT prioritizes deduction and decision-making on the action and interaction of DERs within a complex system.
Specifically, EIoT-DT endeavors to project physical entities (PEs) in the real EIoT onto virtual entities (VEs) in the digital space, contributing to the facilitation of big model (BM) (refer to Section~\ref{Sec:BM}).
The BM, by integrating a system simulator, a situational awareness, and a decision-maker, serves a pivotal role in converting extensive field data and expert experience into informative criteria for domain-specific decision-making, surpassing mere for visualization.
The BM acts as a data-intensive intelligence engine for various downstream domain-specific tasks, with the goal of enhancing management and adding value to environmental PEs through monitoring (understanding the state of EIoT and its DERs), testing (comprehending the reaction to environment changes and its consequences), and optimizing (striving for improved decision-making).

To support such a BM, the VE undergoes an evolution beyond being a mere replica of the real-world PE.
Instead, VE is empowered for ``free-style'' evolution in a ``parallel universe'' or ``mateverse'', \textbf{needing only to follow a few basic laws}.
This evolution of VE enables the generation of multi-scenario (virtual) data for BM training.
In concrete, our research progresses by customizing the VE (image) based on its corresponding PE (inverse image), with an emphasis on DERs' fundamental properties and potential behaviors (refer to Section~\ref{Sec:VEModel}).
Subsequently, Monte Carlo simulations---accounting for various scenarios, multiple time-scales, distinct evolution-paths, etc.---are conducted to generate nearly full-coverage data in parallel universe.
These data facilitate the self-adaptive, self-learning, and self-evolving BM for EIoT at a macro-level.
In such a manner, \textbf{DT provides a natural systematic life-cycle SA} in a virtual space, termed DT-SA.
In general, our DT-SA comprises four key steps---digitalization, simulation, informatization, and intellectualization; the details are elaborated in Section~\ref{Sec:Ingredient}.


The rest of this paper is structured as follows:
Section~\ref{Sec:Background}  delves into the roadmap of DT-SA in EIoT, with an emphasis on the spatial-temporal data utilization (refer to Section~\ref{Sec:STData}).
Section~\ref{Sec:DTRedefinition} presents a redefinition of DT (more like metaverse), along with the designed DT-SA framework, focusing on VE model (refer to Section~\ref{Sec:VEModel}).
Section~\ref{Sec:case} backs DT-SA up with case studies using field data.
Section~\ref{Sec: Conclu} concludes our work.

%

\section{Roadmap of DT-SA in Fourth Paradigm}
\label{Sec:Background}
\begin{figure}[htbp]
\centerline{
\includegraphics[width=.47\textwidth]{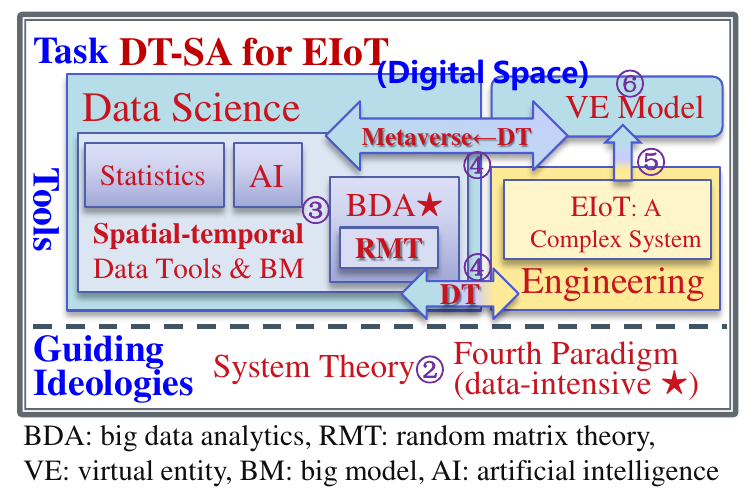}
}
\caption{A brief roadmap of DT-SA for EIoT}
\label{fig:RM4DTSA}
\end{figure}

A concise roadmap of DT-SA, as depicted in Figure~\ref{fig:RM4DTSA}, is presented first to provide an overview of this work.
Rooted in system theory and the Fourth Paradigm, DT-SA is deeply intertwined with state-of-the-art data technologies and disciplines.
In DT-SA framework, DT serves as the data engine, BM as the intelligence engine, and BDA as the core.

\subsection{Background of Twin $\rightarrow$ Digital Twin $\rightarrow$ DT-SA}
The concept of ``twin'' dates back to the 1960s Apollo Program.
In Apollo Program, 15 simulators were created as twins for various tests, preparing for the {unexpected}.
These twins, positioned in an earth-based mode, allowed engineers to {simulate} each condition of the space vehicle (PE) during a flight mission using real-time flight data, thereby assisting astronauts in critical situations.
The twin played a {vital role} in the rescue mission, successfully bringing the astronauts home when disaster struck the Apollo 13 in 1970~\cite{8472083}.

With the development of digital technology~\cite{al2020iot,bouneb2022distributed,kiran2022efficient}, the physical component of the twin had been gradually replaced by its digital counterpart, giving rise to the digital twin (DT).
Initially proposed by Grieves in 2002~\cite{grieves2014digital}, DT gained practical definition and public acknowledgment in NASA's integrated technology roadmap in 2012~\cite{shafto2012dt}.
Various definitions of DT found in the literature are summarized in Table~\ref{tab:Nomenclature}.

\begin{table}[htbp]
\caption{Definitions of DT used in literature}
\label{tab:Nomenclature}
\centering
\begin{tabularx}{0.48\textwidth} { l !{\color{black}\vrule width1pt} p{7.5cm} } 

\toprule[1.5pt]
\hline
\textbf{Ref.} & \textbf{Definition} \\
\Xhline{1pt}
\cite{glaessgen2012digital} & An \textbf{integrated multiphysics, multiscale, probabilistic simulation} of an as-built vehicle or system that uses the best available physical models, sensor updates, fleet history, etc., to mirror the life of its corresponding flying twin. (2012)\\
\cite{shafto2012dt} & An \textbf{integrated multiphysics, multiscale simulation} of a vehicle or system that uses the best available physical models, sensor updates, fleet history, etc., to mirror the life of its corresponding flying twin. DT is \textbf{ultra-realistic} and may consider one or more important and \textbf{interdependent} vehicle system, including propulsion/energy storage, avionics, life support, vehicle structure, thermal management/TPS, etc. Manufacturing anomalies that may affect the vehicle may also be explicitly considered. (2012)\\
\cite{grieves2017digital} & A set of virtual information constructs that \textbf{fully describes} a potential or actual physical manufactured product from the micro atomic level to the macro geometrical level. At its \textbf{optimum}, any information that could be obtained from inspecting a physical manufactured product can be obtained from DT. (2017)\\
\cite{soderberg2017toward} & A digital copy of the physical system to perform \textbf{real-time optimization}. (2017)\\
\cite{Developing2017IRS} & A \textbf{real time digital replica} of a physical device.  (2017)\\
\cite{tao2018digitalDef} & A real mapping of all components in the product \textbf{lifecycle} using physical data, virtual data and interaction data between them.  (2018)\\
\cite{bolton2018customer} & A \textbf{dynamic} virtual representation of a physical object or system across its \textbf{lifecycle}, using \textbf{real-time data} to enable \textbf{understanding, learning and reasoning}.  (2018)\\
\cite{el2018digital}& A digital replica of a living or non-living physical entity. By bridging the physical and the virtual world, data is \textbf{transmitted seamlessly} allowing the virtual entity to exist \textbf{simultaneously with} the physical entity.   (2018)\\
\cite{bolton2018gemini} & A \textbf{realistic digital representation} of assets, processes or system in the built or natural environment.  (2018)\\
\toprule[1pt]
{} & \textbf{Definition in this work}, see Section~\ref{Sec:DTRedefinition} for details\\
\hline
{} & A self-evolving digital organism in the context of {the Fourth Paradigm}, with {massive VEs as molecules/cells/tissues and a built-in BM as brain}. These living VEs are closely linked to their real-world PEs through data, with the goal of  enhancing management and adding value for numerous downstream domain-specific tasks in practice, through monitoring (understanding the state of EIoT units), testing (comprehending the reaction to environment changes and its consequences), and optimizing (striving for improved decision-making) \textbf{in the virtual space}.  (2023)\\
\toprule[1pt]

\end{tabularx}
\end{table}

In the context of the Fourth Paradigm, DT has evolved into a cross-disciplinary product, involving integrants such as data, data technologies/disciplines, domain-specific knowledge, simulation, and analysis.
DT stands out as a promising enabling technology, bridging the gap between data and demands.
With significant advancements in a range of domain-specific tasks, DT has been successfully implemented in numerous engineering fields to date~\cite{Tao2018Digital}, including the field of energy~\cite{he2020opportunities, he2023situation}.
DT market is estimated to grow from USD 3.1 billion in 2020 to USD 48.2 billion by 2026~\cite{markets2020DTmarktet}.

SA, as defined in Section~\ref{sec:motiv}, strives to translate the presented system into the perceived system for decision-making.
Analogous to the Apollo twins (PE--vehicle in orbit, VE--twins on earth) assist astronauts,  \textbf{DT provides a natural systematic DT-SA to help EIoT operators/users in making reliable decisions under normal or urgent conditions}.

\subsection{Challenge of DT-SA: Complexity, Emergence Phenomenon}
\label{Sec: motivationofAgg}

\textbf{Our exploration of DT-SA is primarily motivated by comprehending the intricate emergence phenomena within complex EIoT}.
This motivation is anchored in our quest to harness the positive emergence, as it plays a pivotal role in the effective functioning of EIoT.

\subsubsection{Aggregation behaviors and its emergence phenomena}
\label{sec:AggandEmg}
\text{\\}

\begin{figure*}[htbp]
\centerline{
\includegraphics[width=.88\textwidth]{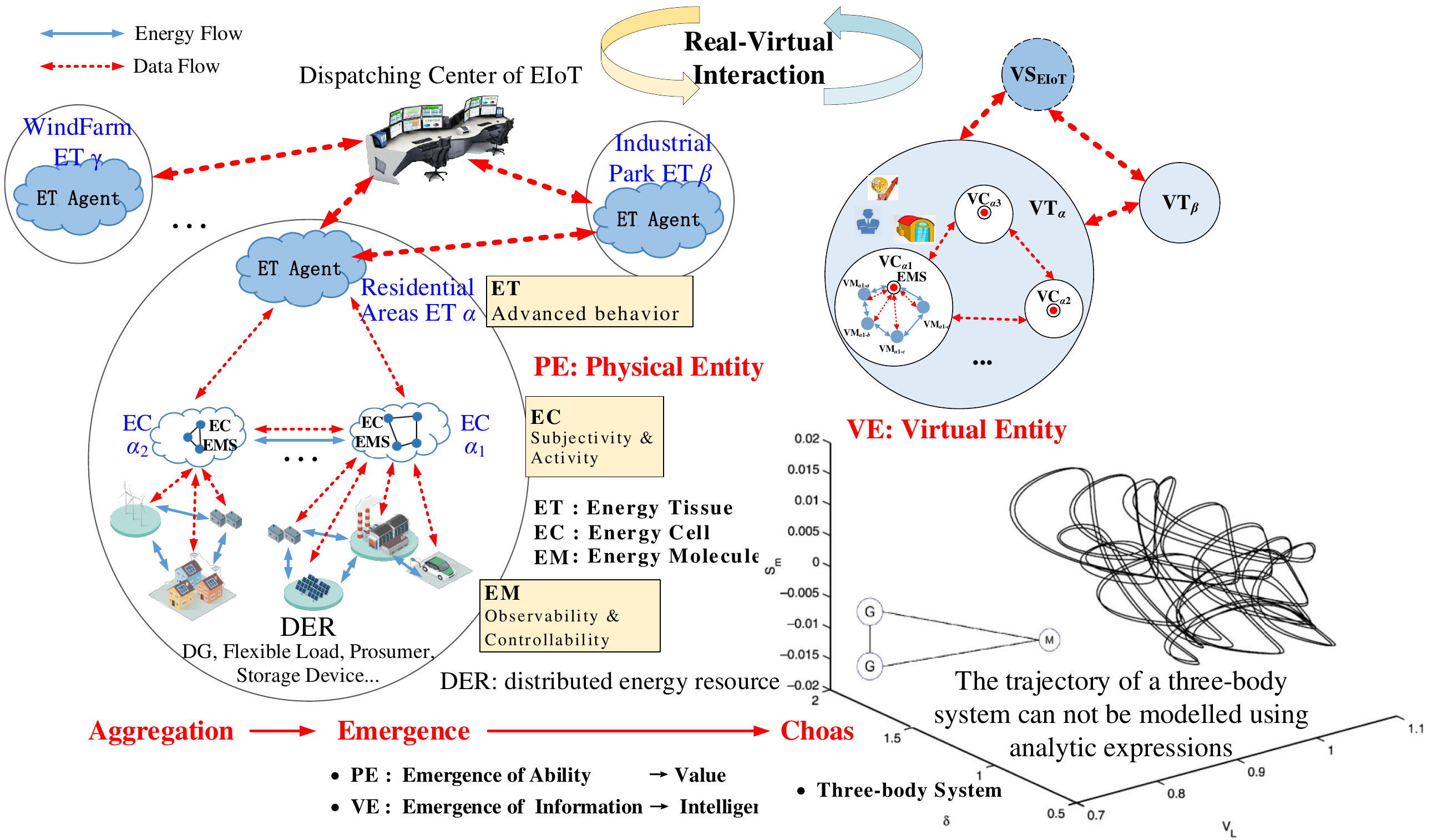}
}
\caption{Motivation and engineering background of DT-SA:
Aggregation behavior and its emergence phenomena in EIoT. }
\label{fig:EIoTAgg}
\end{figure*}

A standard EIoT at the distribution network level is depicted in Figure~\ref{fig:EIoTAgg}.
In essence, EIoT~\cite{gupta2020overview} is a network connecting DERs, characterized by their small size, large number, distributed deployment, hierarchical structure, diverse behavior, smart response, and uncertain control.
To enhance management efficiency, aggregation behaviors are implemented to those DERs, leading to a hierarchical complex EIoT---these DERs converge to create an aggregation termed ``energy cell'' (e.g., intelligent building, microgrid), and furthermore form some advanced aggregation termed ``energy tissue''(e.g., microgrid cluster, virtual power plant).

\begin{figure}[htbp]
\centerline{
\includegraphics[width=.48\textwidth]{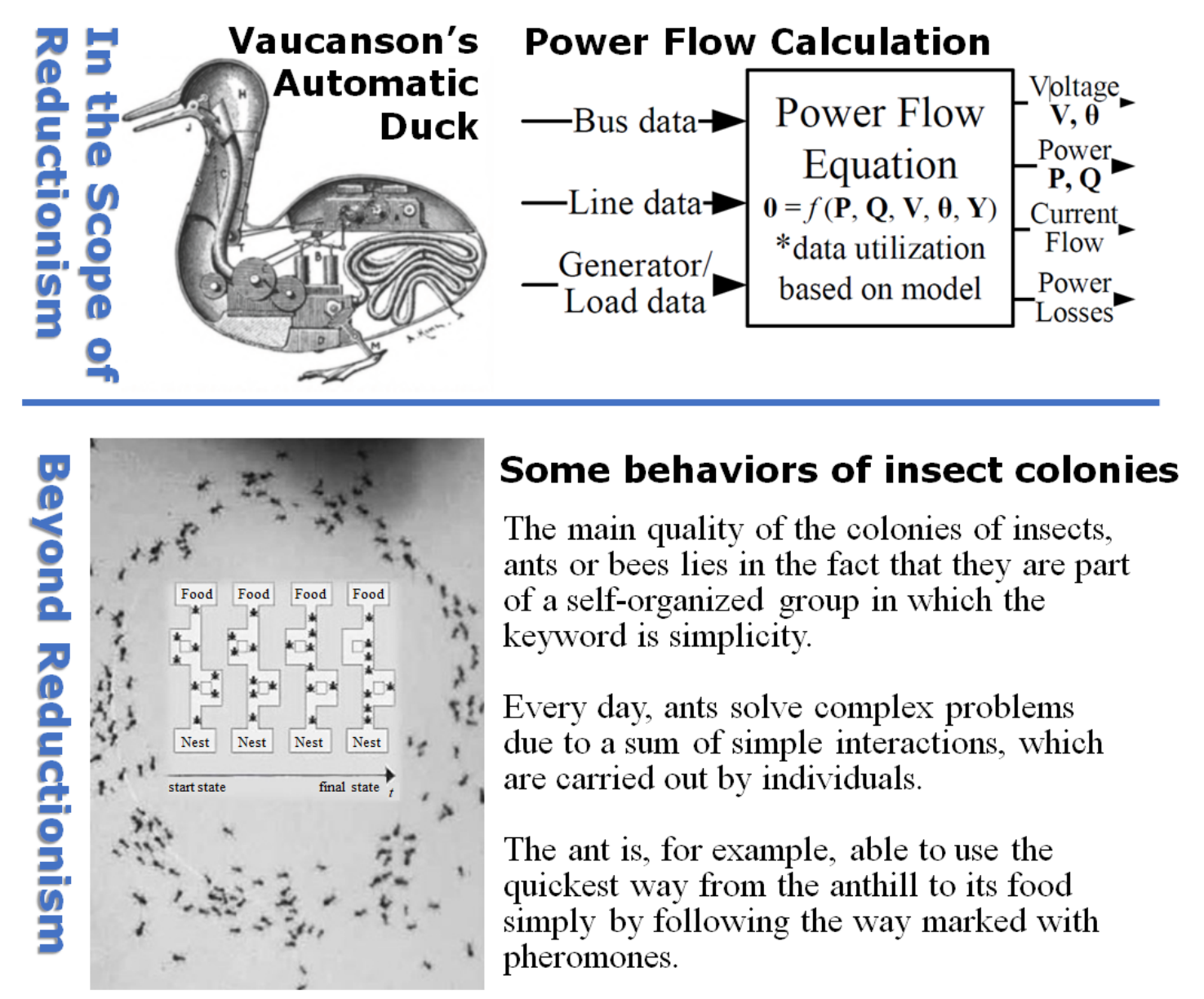}
}
\caption{Schema of emergence phenomena and reductionism.}
\label{fig:Redu}
\end{figure}

Aggregation behaviors may give rise to emergence phenomena, either positive or negative.
Emergence occurs when an entity exhibits traits that its components do not possess individually---in other words, the whole is more than the sum of its parts.
Emergence phenomena are common in real-life \textbf{but typically beyond reductionism}, as depicted in Figure~\ref{fig:Redu}.

It is paramount to emphasize that not every aggregation behavior or complexity leads to emergence phenomena.
Certain instances, such as Vaucanson's automatic duck and power flow calculation, fall in the scope of reductionism.
Despite their precision and complexity, their intricate patterns can be methodically modeled piece-by-piece, yielding analytical formulations in the form of $\bm x\!=\!f(t)$.
In contrast, reductionism proves inadequate when applied to some behavior exhibited by insect colonies.
\textbf{A holistic perspective, beyond reductionism, becomes imperative in such cases, as these behaviors are indivisible}. For instance, ants construct fancy underground palaces through a rather simple cooperative mode without a super-brain conductor; for each individual ant, it is impossible to formulate its trajectories using an analytical description in the form of $\bm x\!=\!f(t)$.

%
%

Back to EIoT as depicted in Figure~\ref{fig:EIoTAgg}, user-side DERs are aggregated aiming to induce positive emergence: a) the emergence of intelligence, fostering a smarter decision-maker, and b) the emergence of value, enhancing operational efficiency.

\subsubsection{Chaos, Lorenz system, and multi-body system}
\label{sec:chaosuniver}
\text{\\}

It is crucial to acknowledge that \textbf{the aggregation behaviors may inevitably result in negative emergence phenomena}, such as chaos, a special outcome of complexity that is \textbf{beyond the scope of reductionism}.
To quickly grasp the concept of chaos, consider a classical chaotic system, the Lorenz system, represented as Eq.~\eqref{eq:Lorenz}:
\begin{equation}
\label {eq:Lorenz}
\begin{aligned}
\dot{x}_1 & =\sigma({x}_2-{x}_1) \\
\dot{x}_2 & =\rho {x}_1-{x}_2-{x}_1 {x}_3 \\
\dot{x}_3 & =-\beta {x}_3+{x}_1 {x}_2
\end{aligned}
\end{equation}
Mathematically, an analytical solution in the form of $\bm x\!=\!f(t)$ cannot be deduced from Eq.~\eqref{eq:Lorenz}, classifying the Lorenz system as a ``chaotic system''. Further details about this experiment can be found in Reference~\cite{moon2017periodicity}.

Similar to the Lorenz system, a three-body system may also exhibit chaotic behavior, as illustrated in the bottom-right part of Figure~\ref{fig:EIoTAgg}.
In EIoT, a multi-body system with chaotic effects,
\textbf{the intricate interactions among DERs or their aggregations may induce unintended consequences or even serious incidents}.
Therefore, a profound understanding of both positive and negative emergence is pivotal for the effective management of DERs, contributing to advanced decision-making within the complex EIoT landscape.

\subsubsection{Chaos in EIoT and ideas for solutions}
\label{sec:chaos4EIoT}
\text{\\}

\begin{figure}[htbp]
\centering
\subfloat[Chaotic power system model with three buses]{\label{chaos-0}
\includegraphics[width=0.40\textwidth]{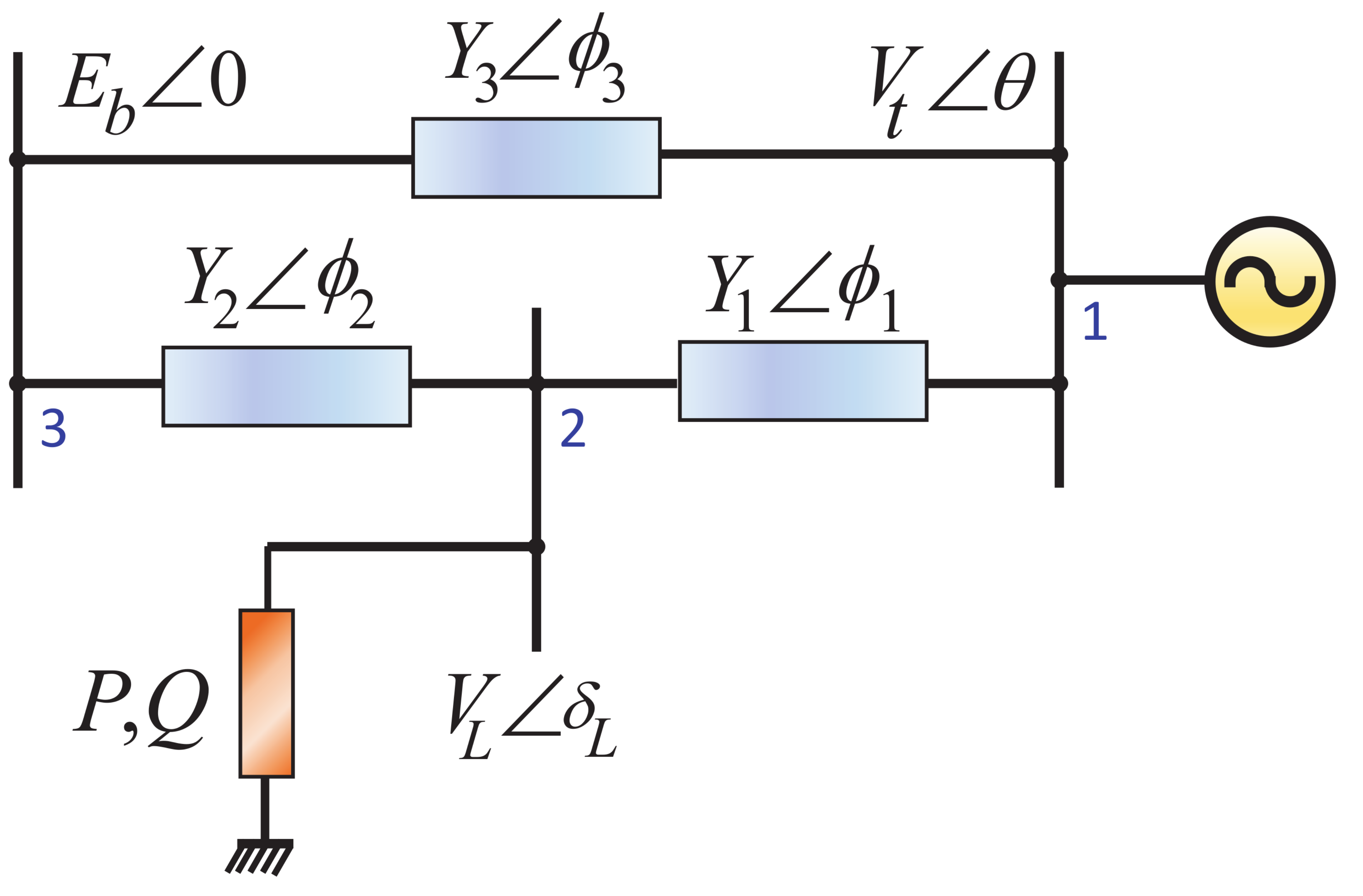}}

\subfloat[Time-series waveforms]{\label{chaos-a}
\includegraphics[width=0.40\textwidth]{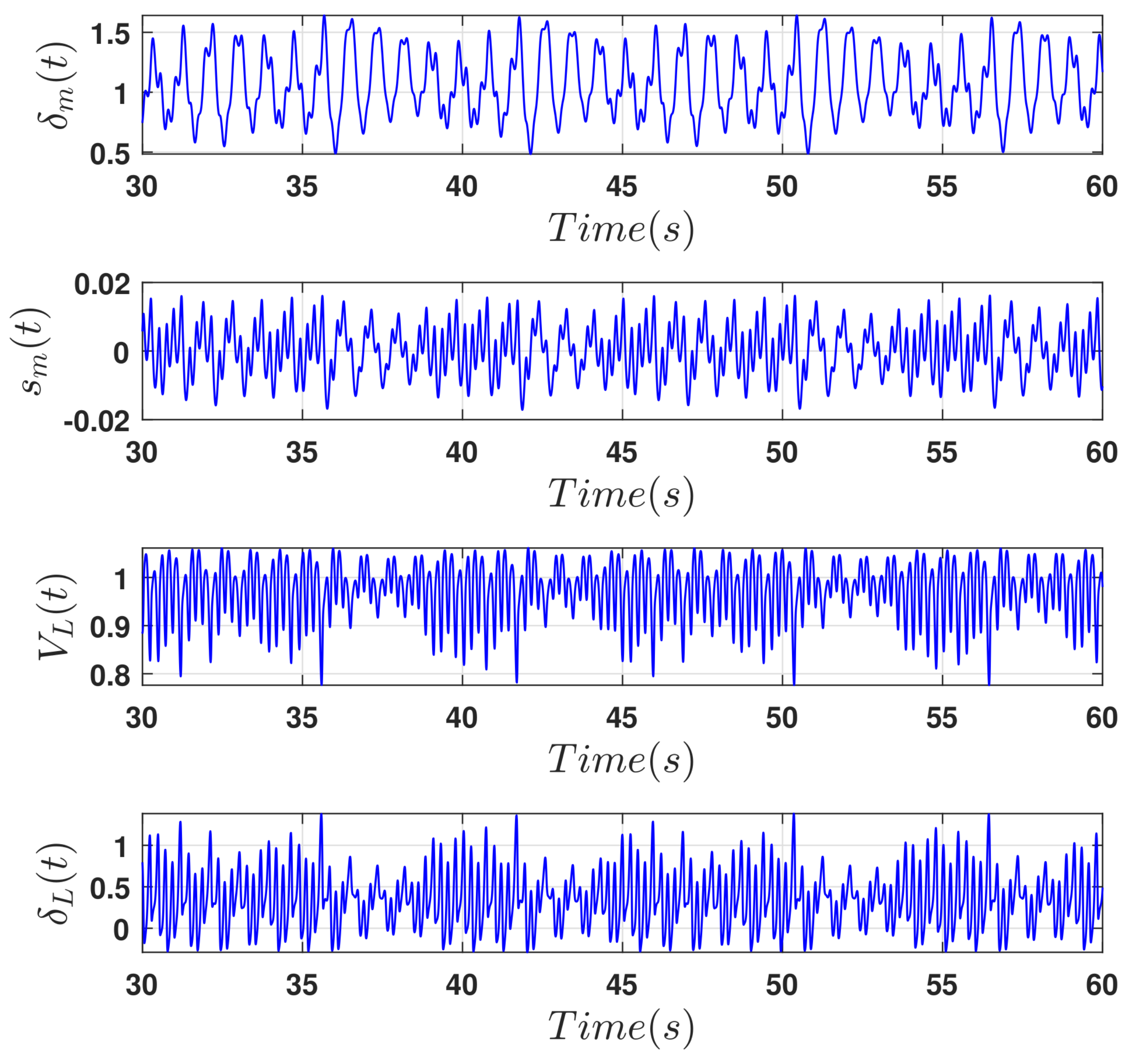}}

\subfloat[Phase portrait strange attractor]{\label{chaos-b}
\includegraphics[width=0.40\textwidth]{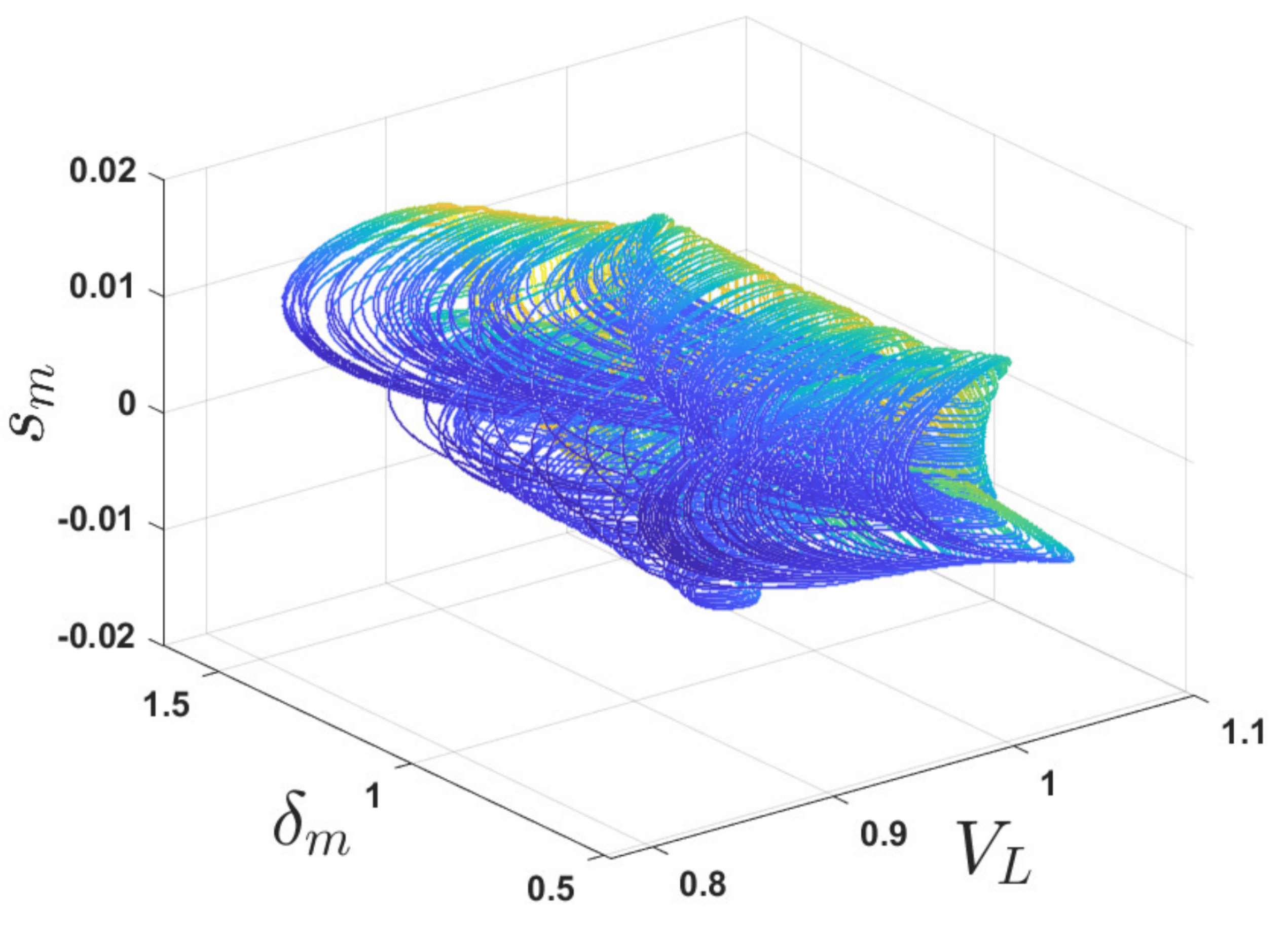}}

\caption{The chaotic power system~\cite{electronics10131532}. }
\label{fig:ChaosSystem}
\end{figure}

In the domain of power systems, a chaotic system can be triggered when a single generator bus (Bus 1) supplies power to a local load bus (Bus 2) and connects to a slack bus (Bus 3), as depicted in Figure~\ref{chaos-0}~\cite{electronics10131532}.

The dynamics of the system depend on various parameters ($\delta_m$--generator angle, $s_m$--generator slip, $E'_q$--generator q-axis transient potential , $E_{fd}$--excitation potential, $\delta_L$--load phase angle, and $V_L$--load bus voltage amplitude), as expressed in Eq.~\eqref{eq:choasSyS}.
Figure~\ref{chaos-a} and~\ref{chaos-b}, respectively, illustrate the time series waveforms of each variable and the phase portrait strange attractor in 3-dimensional space, confirming the chaotic nature of the power system.
Details of this chaotic power system can be found in Reference~\cite{electronics10131532}.

\begin{equation}
\label{eq:choasSyS}
\left\{\begin{array}{l}
\dot{\delta}_m=\omega_B s_m, \\
\dot{s}_m=\frac{1}{2 H}\left(-d s_m+P_m-P_g\right) \\
\dot{E}_q^{\prime}=\frac{1}{T_{d 0}^{\prime}}\left(-E_q^{\prime}+\left(x_d-x_d^{\prime}\right) I_d+E_{f d}\right), \\
\dot{E}_{f d}=\frac{1}{T_A}\left(-E_{f d}+K_A\left(V_{r e f}-V_t\right)\right), \\
\dot{\delta}_L=\frac{1}{q_1}\left(Q-Q_{1 d}-Q_0-q_2 V_L-\left(q_3-B_c\right) V_L^2\right) \\
\dot{V}_L=\frac{1}{p_2}\left(P-P_{1 d}-P_0-p_3 V_L-{p_1}\dot{\delta}_L\right).
\end{array}\right.
\end{equation}

In analyzing such a system,  we should set aside reductionism, acknowledging that {the chaotic trajectory cannot be analytically formulated}.
Fortunately, while chaos and emergence may appear random, they inherently follow some patterns hidden in the spatial-temporal data, such as the time-series waveforms in Figure~\ref{chaos-b}.
This viewpoint will be exemplified through a case (see Section~\ref{sec:matrix4choas}) only, and will not be further explored in this paper.
Moreover, the emergence phenomenon, usually composed of numerous simple behaviors, is relatively easy to reproduce through simulations.
Therefore, for the analysis of a chaotic EIoT with aggregation behaviors and emergence phenomena, we endorse \textbf{data-intensive solutions}, with an emphasis on data-driven mode, systematic analysis, and numerical expression.
In this context, \textbf{DT technology sets a solid cornerstone for gaining insight into EIoT.}

\subsection{Guiding Ideologies: Fourth Paradigm, System Theory}

The scientific paradigm continues its evolution in history: starting with empirical science focused on direct observation, transitioning to theoretical science with mathematical models, advancing to computational science using simulations for complex phenomena, and culminating in today's eScience, which integrates theory, experiment, and simulation through data exploration, leveraging large datasets and advanced data management techniques.
Presently, with the influx of heterogeneous data from sensors/simulations, all scientific paradigms---experimental, theoretical, and computational---are influenced by this data deluge.
\textbf{Insights from the data are gained farily late} in the data-processing pipeline, marking a paradigm shift towards a data-intensive scientific discovery.
Today's scientific paradigm, along with its tools, techniques \& technologies, is so different that it deserves distinguishing data-intensive science from computational science as a new, the Fourth Paradigm for data-intensive scientific discovery~\cite{hey2009fourth}.

In the context of the Fourth Paradigm, our approach involves employing system theory and statistics to tackle complexity.
General system theory, introduced by Bertalanffy in 1950~\cite{von1950outline},  aims to establish universal principles applicable to systems regardless of their specific elements and forces.
Nothing prescribes that we have to end with analytically results.
In fact, this view was not altered but rather reinforced when deterministic laws were gradually replaced by statistical laws~\cite{von1973meaning}.

Starting from original general system theory, the term complex adaptive system (CAS) was coined by sociologist Buckley in 1968~\cite{buckley1968modern}.
Holland further developed CAS, defining it as ``systems that have a large number of components, often called agents, that adapt or learn as they interact''~\cite{holland1992complex}, with an emphasis on system's \textbf{complexity, emergence, and other macroscopic properties}~\cite{holland2000emergence}.
Our work extends the concept of agent into molecule--cell--tissue--system, \textbf{reflecting the hierarchical structure with aggregation behavior of DERs in complex EIoT} (refer to Figure~\ref{fig:EIoTAgg}). This extension influences the design of our unified Virtual Entity (VE) model (refer to Section~\ref{Sec:VEModel}).

\subsection{Big Model for Complex System}
\label{Sec:BM}
Human behaviors inspire the concept of the big model (BM). We, as human beings,
start our lives by learning a wide range of knowledge, without domain-specific problem-solving, to form our initial outlook.
This initial outlook has a potential influence on our decision-making in many aspects as we grow up.
In light of this, the traditional model construction pattern ``different models for different tasks'' is gradually being replaced by ``one data-intensive pre-built BM for various tasks''---researchers collect massive data and develop advanced algorithms, in order to build a BM for a large-scale system with different demands~\cite{yuan2022roadmap}. From this point, \textbf{BM is a data-intensive by-product of Fourth Paradigm}.

The BM functions as a built-in intelligence engine that absorbs the laws and patterns hidden in data.
Its utilization significantly reduces the reliance of downstream tasks on vast amounts of high-quality data.
As a result, scenarios that were previously challenging due to limited labeled data become more tractable.
This paradigm shift can be compared to our power system in particular---BM serves as a "slack bus," generating high-quality intelligence and thus serving a number of downstream applications.

%
%

There are mainly three types of models in the built-in BM:
a) white box---mechanism model; b) black box---deep neural network model; c) gray box---statistical model.
It's imperative to underscore that the preceding sections (Section~\ref{sec:motiv}, \ref{Sec: motivationofAgg}) have emphatically conveyed that relying solely on either white box or black box models falls short in effectively addressing emergence and chaos phenomena within the EIoT. \textbf{As a natural progression, our focus now shifts to the gray box}, specifically the high-dimensional analysis in Section~\ref{Sec:STData}. Furthermore, we delve into BM's ecosystem, DT, in Section~\ref{Sec:DTRedefinition}.

\section{Spatial-temporal Analysis for DT-SA}
\label{Sec:STData}

\begin{figure*}[htbp]
\centerline{
\includegraphics[width=.88\textwidth]{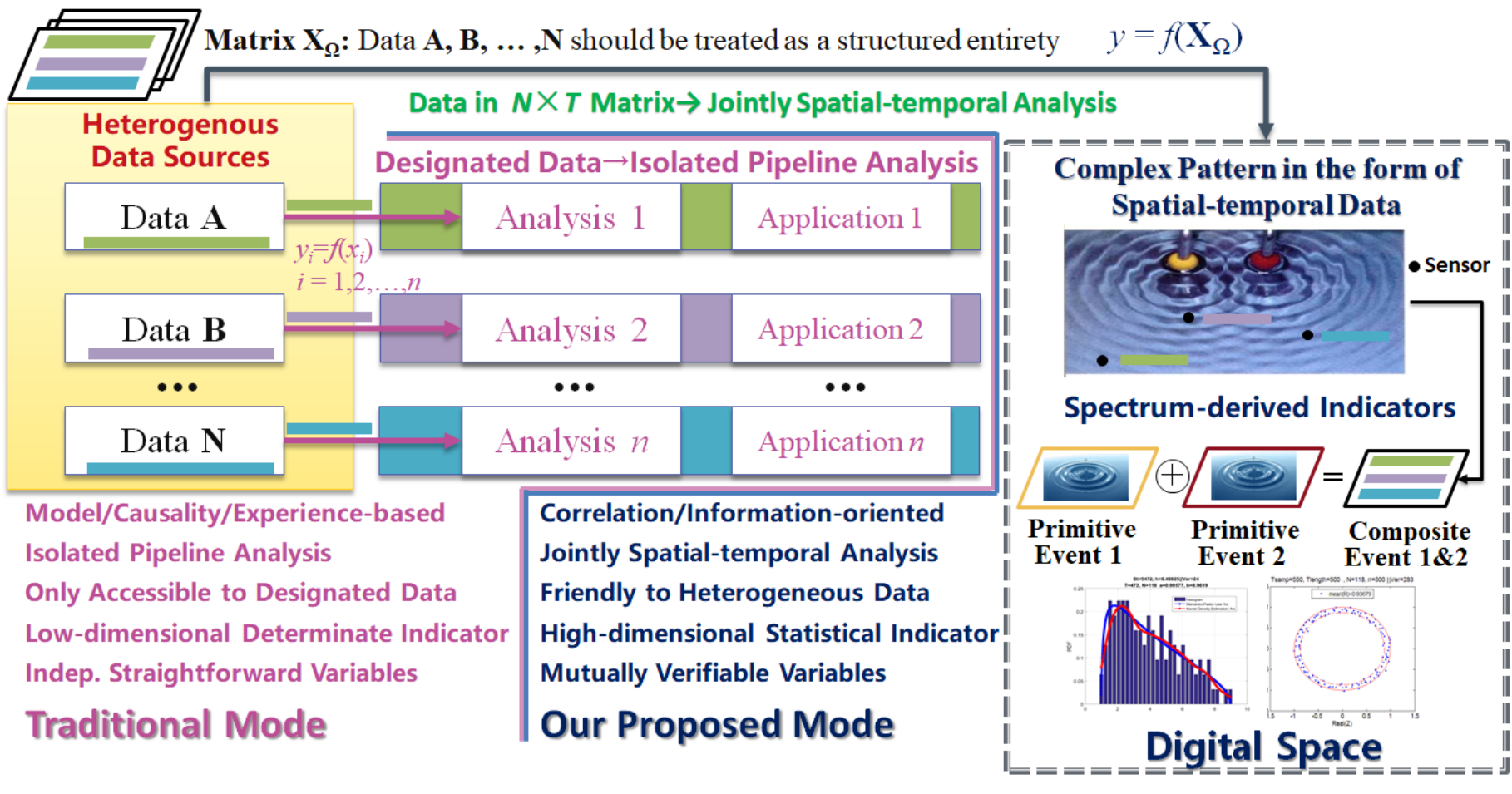}
}
\caption{Independently spatial-temporal analysis paradigm V.S. jointly spatial-temporal analysis paradigm}
\label{fig:DTVSModel}
\end{figure*}

As previously discussed, spatial-temporal analysis holds paramount importance within the realms of DT-SA and BM.
However, a significant portion of existing analyses, primarily mechanism-based and white-box, tends to be carried out in an isolated (spatial-temporally disjointed) manner within a low-dimensional space, as depicted in the central segment of Figure~\ref{fig:DTVSModel}.
Unfortunately, these analyses frequently grapple with limitations, as they are predominantly geared towards addressing ideal, typical, or extreme scenarios.
In light of these constraints, we embark on a comprehensive exploration of spatial-temporal analysis for DT-SA.

\subsection{Spatial-temporal Data and its Mining}
\label{Sec:Datautilization}

As emphasized earlier, spatial-temporal analysis plays a pivotal role in the context of DT-SA and BM.
In spatial-temporal analysis, we routinely handle multiple variables (spatial space, $N$), where each variable ($i\!=\!1,..., N$) samples time-series data for a given duration (temporal space, $T$) of observation. For instance, the time-series waveforms, as illustrated in Figure~\ref{chaos-a}, represent the chaotic behavior of the power system (Figure~\ref{chaos-0}).

Traditional statistical theory usually deals with a fixed $N$ (commonly $N$ is small, typically $N\!<\!6$~\cite{qiu2015smart}). For instance, for Part Transformation (ABC--DQ0, in the power system domain), $N\!=\!3$.
This fixed small $N$ is referred to as the low-dimensional regime.
In engineering scenarios, however, we are often interested in cases where $N$ can \textbf{vary arbitrarily in size compared to} $T$  (commonly $T$ is large, typically $T\!>\!60$, $N\!>\!10$, $N/T\!>\!0$~\cite{qiu2015smart}).
This fundamental requirement serves as the \textbf{primary driving force} for our exploration of BDA (big data analytics).
Consequently, we adopt a (large-dimensional) data matrix, rather than a vector or a scalar, as the foundation of BDA.
These data matrices can be viewed as data points in high-dimensional vector space.

For DT-SA in EIoT, (spatial-temporal) data mining is expected to yield high-dimensional information with domain-specific significance.
The data resource (\emph{input}) primarily comprises sampled spatial-temporal data denoted as $\mathbf X\!\in\!\mathbb{R}^{N\!\times \!T}$.
Concerning high-dimensional information (\emph{output}), there are mainly two types: a) High-dimensional indicators for high-dimensional statistics, and b) deep features for deep learning.

Once the input and output are established, attention turns to information extraction tools.
Traditional mathematical tools often prove inadequate for handling spatial-temporal data~\cite{qiu2015smart}.
This inadequacy has greatly spurred the development of data science, particularly in the fields of AI (artificial intelligence) and BDA. Accordingly, we highlight a focal tool for each field: a) AI field---deep learning (DL), which does well in massive data modeling, and b) BDA field---High-dimensional statistics, specifically random matrix theory (RMT), which is good at spatial-temporal data analytics.
Both tools \textbf{involve a series of high-dimensional methods for jointly spatial-temporal modeling and analysis}, and have already \textbf{made profound impacts in various fields}.

\subsubsection{Deep Learning and the Advantages}
\label{sec:DLofAI}
\text{\\}

DL stands as the {state-of-the-art} data mining algorithm. It aims to extract high-level, complex abstractions (i.e. deep features) through a hierarchical learning process. In this process, deep features are learnt at a given level based on relatively lower features (i.e., the features in the lower layer of the network)~\cite{najafabadi2015deep}.
Throughout this learning process, DL utilizes extensive data, represented by $\mathbf{x}$, in a non-handcrafted manner, facilitating the construction of a deep network model
\[  \mathbf{y}\!=\!{{\bm f}^{L}}\left( {{\mathbf{W}}^{L}}  \!\cdots\! {{\bm f}^{2}}\left( {{\mathbf{W}}^{\text{2}}}{{\bm f}^{1}}\left( {{\mathbf{W}}^{1}}\mathbf{x}\!+\!{{\mathbf{b}}^{1}} \right)\!+\!{{\mathbf{b}}^{\text{2}}} \right)\!\cdots\! \!+\!{{\mathbf{b}}^{L}} \right)
\]

DL models are inherently data-driven, allowing them to \textbf{adaptively adjust their parameters with little prior knowledge} of the physical mechanism or causal relationship. Consequently, DL can be generalized to different scenarios or even different systems without significant modifications.

\subsubsection{Big Data Analytics \& RMT and the Advantages}
\label{sec:RMTofBDA}
\text{\\}

BDA aims to extract valuable insights (in the form of high-dimensional statistics) from {matrix-based variables} (such as eigenvalues or the matrix variate itself~\cite{adhikari2007matrix}) through {jointly temporal-spatial analysis}.
Handling data that is \textbf{high in dimensionality}, rather than large in size, poses challenges in making BDA {tractable}.
The theoretical aspects are discussed in more detail in the next subsection, Subsection~\ref{sec:BADandRMT1}, while case studies are presented in Subsections~\ref{sec:matrix4choas} and~\ref{sec:Case1}.

\subsection{Random Matrix Theory and its LES Indicator for BDA}
\label{sec:BADandRMT1}
Given the scarcity of labeled data in engineering scenarios, especially data annoteated with ``abnormal operating condition'' labels, there is  \textbf{significant potential for the exploration and effectiveness of unsupervised algorithms}. In this regard, RMT stands out as a noteworthy unsupervised algorithm.
RMT has recently become analytically tractable~\cite{qiu2015smart}, offering transparency and rigorous handling of data matrices.

The primary goal of RMT is to comprehend the joint eigenvalue distribution, which serves as statistical analytics for data matrices in the asymptotic regime---high-dimensional analysis and visualization are treated as functionals of eigenvalue distributions.
For example, linear eigenvalue statistics (LES)\cite{shcherbina2011central} of the matrix exhibit Gaussian distributions under very general conditions.
Recent breakthroughs in probability, particularly the Central Limit Theorem (CLT) for the LES\cite{qiu2015smart}, have allowed the study of other statistical variables, making their properties derivable and provable.
In this sense, RMT is much more fundamental than DL.
Besides, RMT excels in handling {moderate-size (unlabeled)} data, a common scenario in the EIoT landscape.
For the sake of completeness of this paper, we provide a brief expansion on the LES. Additional details can be found in~\cite{he2015arch}

\subsubsection{Definition of LES}
{\text{\\}}

Consider a random matrix $\mathbf{\Gamma}\!\in\! {{\mathbb{R}}^{N\!\times\! T}}$, and  $\mathbf{M}$ is the covariance matrix ${{\mathbf M}}\!=\!{\frac {1}{N}\mathbf{\Gamma}\mathbf{\Gamma}^{\mathrm H}}$.
The LES $\tau$ of  $\mathbf{\Gamma}$ is defined in \cite{lytova2009clrforles,shcherbina2011central}
 via the continuous test function  ${\varphi: \mathbb C \rightarrow \mathbb C,}$
\begin{equation}
\label{eq:DDLES}
\tau_\varphi=\sum_{i=1}^{N}{\varphi({\lambda_i})}=\text{Tr}\varphi \left( \mathbf{M} \right),
\end{equation}
where the trace of the function of a random matrix is involved.


\subsubsection{Law of Large Numbers for LES}
{\text{\\}}

The Law of Large Numbers tells us that $N^{-1}\tau_\varphi$ converges in probability to the limit
\begin{equation}
\label{eq:LES1}
\lim_{N \to \infty}\frac{1}{N}\sum_{i=1}^{N}{\varphi({\lambda_i})}\!=\!\int\varphi(\lambda)\rho(\lambda)\,d\lambda,
\end{equation}
{where $\rho(\lambda)$ is the probability density function of the eigenvalue $\lambda$. Therefore, we deduce that}\normalsize{}
\begin{equation}
\label{eq:LES10}
\mathbb{E}(\tau_\varphi)=N\!\int\varphi(\lambda)\rho(\lambda)\,d\lambda,
\end{equation}

\subsubsection{Central Limit Theorem for LES}
\label{hearttheory}
{\text{\\}}

As the natural second step, CLT aims to study the variations of LES $\tau_\varphi$. CLT tells that

\newtheorem{thm55}{Theorem}[section]
\begin{thm55}[M. Sheherbina, 2009, \cite{shcherbina2011central}]
 Let the real valued test function $\varphi$ satisfy condition ${{\left\| \varphi  \right\|}_{3/2+\varepsilon }}\!<\!\infty   \left( \varepsilon >0 \right)$. Then $\!\tau_\varphi\!$, as defined in Eq.~\eqref{eq:DDLES}, in the limit $N,T\!\to\!\infty , {c}\!=\!{N/T}\le 1$, converges in the distribution to the Gaussian random variable with the mean $\mathbb{E}(\tau_\varphi)$, according to Eq.~\eqref{eq:LES10}, and the variance:
\begin{equation}
\label {eq:CLTforLes}
\begin{aligned}
     \sigma^2(\tau_{\varphi})=    &    \frac{2}{c\pi^2 }\iint\limits_{-\frac{\pi }{2}<{{\theta }_{1}},{{\theta }_{2}}<\frac{\pi }{2}}{{{\psi }^{2}}\left( {{\theta }_{1}},{{\theta }_{2}} \right)}\left( 1-\sin {{\theta }_{1}}\sin {{\theta }_{2}} \right) d{{\theta }_{1}}d{{\theta }_{2}} \\
 &                +\frac{{{\kappa}_{4}}}{{\pi }^{2}}\left( \int_{-\frac{\pi }{2}}^{\frac{\pi }{2}}{\varphi \left( \zeta \left( \theta  \right) \right)\sin \theta  d{{\theta }}} \right)^2, \\
\end{aligned}
\end{equation}
{where $\psi \left( {{\theta }_{1}},{{\theta }_{2}} \right)\!=\!\frac{\left[ \varphi \left( \zeta \left( \theta  \right) \right) \right]\arrowvert_{\theta ={{\theta }_{2}}}^{\theta ={{\theta }_{1}}}}{\left[ \zeta \left( \theta  \right) \right]\arrowvert_{\theta ={{\theta }_{2}}}^{\theta ={{\theta }_{1}}}},$
${\left[ {\zeta \left( \theta  \right)} \right]\arrowvert_{\theta  = {\theta _2}}^{\theta  = {\theta _1}}}\!=\!\zeta \left( {{\theta _1}} \right)\! -\! \zeta \left( {{\theta _2}} \right),$
and $\zeta \left( \theta  \right) \!= \!1\! + \!1/c \!+\! {2}\!/\!{\sqrt c} \sin \theta;$   $\kappa_4\!=\!\mathbb{E}\left( {{X_{ij}}^{4}} \right)\! -\!3$ is the $4$-th cumulant of entries of $\mathbf X$.}

\normalsize{}
\label{th555}
\end{thm55}
\normalsize{}

\subsection{LES for Insight into Chaotic Systems}
\label{sec:matrix4choas}
\begin{figure}[htbp]
\centerline{
\includegraphics[width=.49\textwidth]{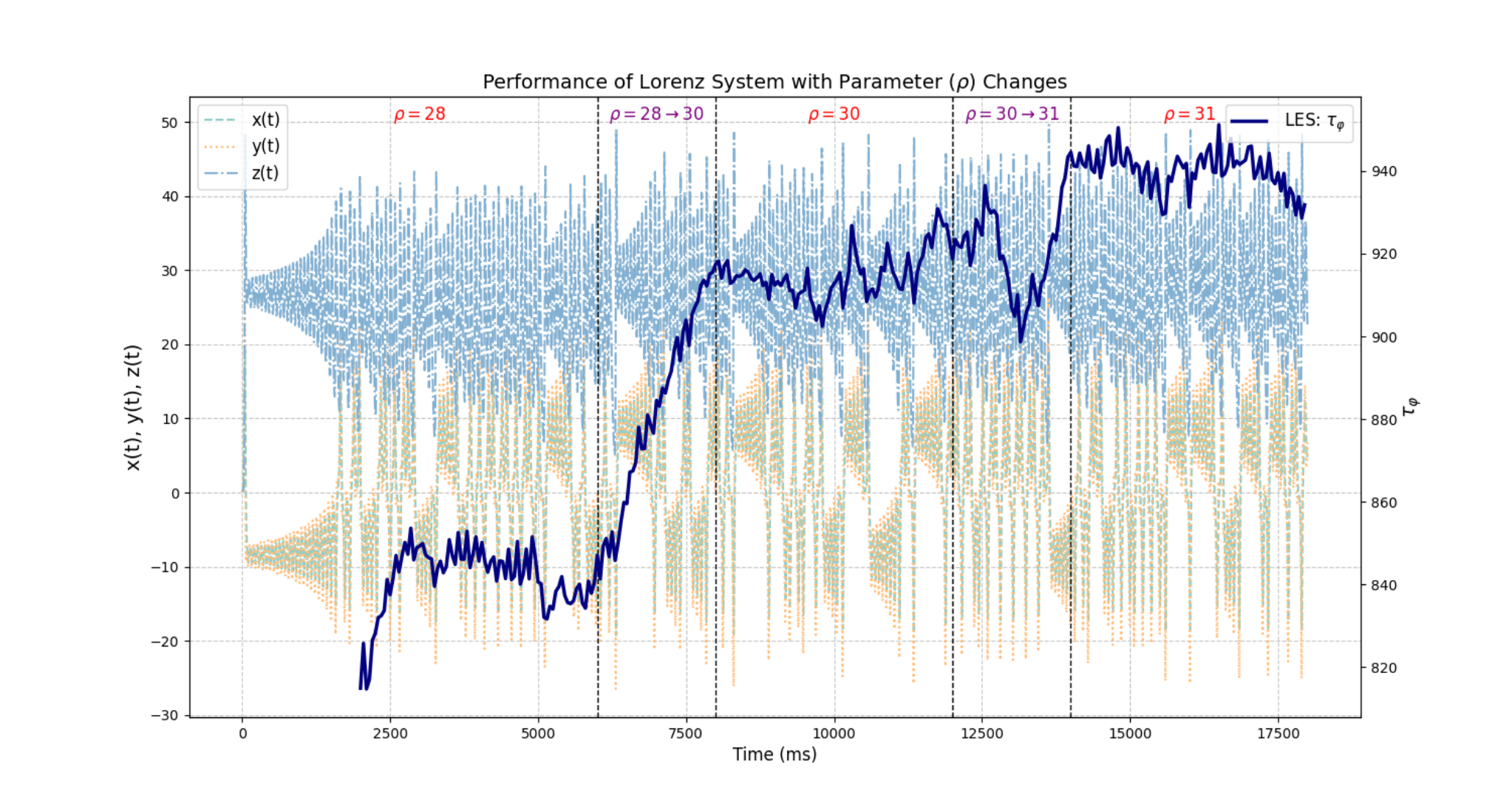}
}
\caption{Performance of Lorenz system with parameter changes}
\label{fig:ChaosLorenz}
\end{figure}

Let's consider the Lorenz system expressed in Eq.~\eqref{eq:Lorenz}, with parameters $\sigma\!=\!10, \beta\!=\!8/3$, and $\rho\!=\!24$.
Then we initiate the simulation of the chaotic system with the following steps:
\begin{enumerate}[a)]
\item Set the simulation frequency to 100 Hz.
\item Introduce change points (CPs) of parameter $\rho$. 1) CP1: $28\!\rightarrow\! 30$ at 60s; and 2) CP2: $30\!\rightarrow\! 31$ at 120s.
\item Generate simulation sequences with a total length of 18000 (3 minutes, $18000\! =\! 3\! \times \!60\! \times \!100$).
\item Employ a moving sliding window (MSW) of length 2000 to sample spatial-temporal data matrix ($X\!\in\! \mathbb{R}^{3\!\times\!2000}$) from the sequence, and then calculate LES indicator $\tau_\varphi$ for $X$.
\end{enumerate}
The performance is obtain as illustrated in Figure~\ref{fig:ChaosLorenz}.

When the parameter $\rho$ undergoes a slight jump at both change points (CP1: at 60s, $\rho\!: 28\!\rightarrow \!30$, CP1: at 120s, $\rho\!: 30\!\rightarrow\! 31$), the raw data appear insensitive. However, employing LES $\tau_\varphi$ allows us to easily distinguish their different patterns in the high-dimensional space.

This case serves as a practical illustration of applying the LES indicator to gain insights into chaotic systems, aligning with the Fourth Paradigm's emphasis on data-intensive scientific discovery.
In this context, \textbf{spatial-temporal data analysis, specifically RMT,  severs as a powerful tool for DT-SA}---it enables the extraction of informative patterns and insights from the extensive data within the complex EIoT.

\section{Redefinition of DT toward Metaverse}
\label{Sec:DTRedefinition}
This section presents a cutting-edge redefinition of DT, aimed at seamless integration with the Fourth Paradigm, system theory, and BM.

\subsection{Redefinition of Digital Twin: Extend DT into Metaverse}
As discussed in Section~\ref{Sec:Background}, today's EIoT urgently demands an advanced DT that surpasses the limitation of being merely an 1-1 replica of its physical counterpart.
In response to this demand, we propose a redefinition of DT to align it with the Fourth Paradigm, CAS, and BM:
\emph{DT is a self-evolving digital organism, with {massive VEs as molecules/cells/tissues and a built-in data-intensive BM as the brain}. These living VEs are closely linked to their real-world PEs through data, with the goal of enhancing management and adding value for {numerous downstream domain-specific tasks} in practice, through monitoring (understanding the state of PEs), testing (comprehending the reaction to environment changes and its consequences), and optimizing  (striving for improved decision-making) {in the virtual space}.}

This redefined DT signifies a departure from the original concept of a 1-1 mirror world. Instead, it advances towards \textbf{digital parallel universes, enabling a comprehensive and holistic consideration in the landscape of complex EIoT}. In this context, \textbf{our redefined DT aligns more closely with the concept of the metaverse}, an emerging paradigm finding extensive applications in various fields such as gaming, military, and sociology. Importantly, both the original DT and the metaverse adhere to IEEE 2888 Standards~\cite{9480910}, with the metaverse expanding into a broader and more intricate scope.

\begin{figure}[htbp]
\centerline{
\includegraphics[width=.48\textwidth]{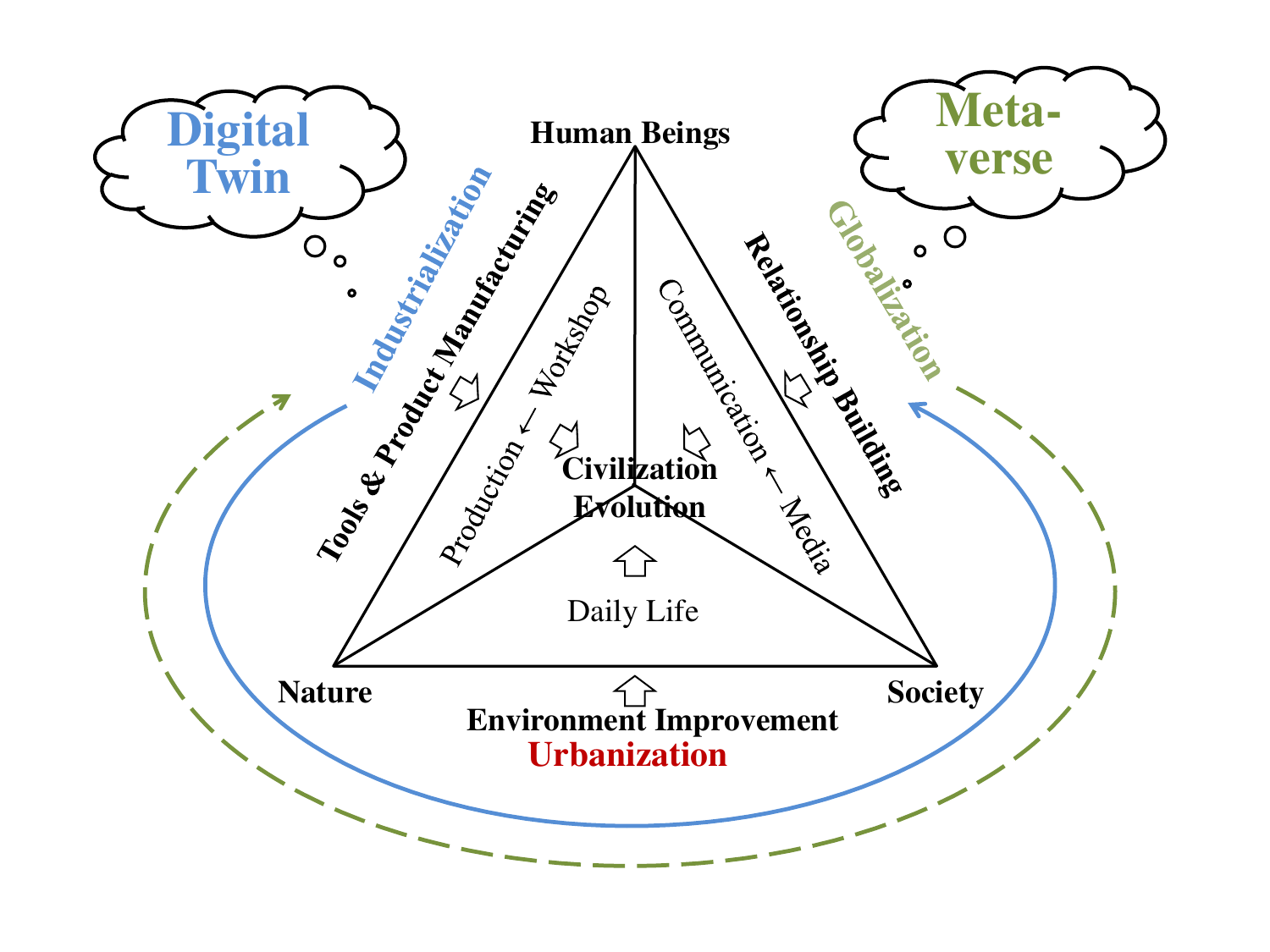}
}
\caption{Metaverse V.S. DT: A technological evolution view~\cite{duan2018meta}}
\label{fig:Metaverse}
\end{figure}

This redefined DT marks a departure from the original DT, a 1-1 mirror world.
Instead, it advances towards \textbf{digital parallel universes, allowing for a comprehensive and holistic consideration in the landscape of complex EIoT}.
In this sense, \textbf{the redefined DT is more akin to metaverse}, an emerging concept gaining extensive applications in various fields such as gaming, military, sociology.
Notably, both the original DT and the metaverse conform to IEEE 2888 Standards~\cite{9480910}, with the metaverse expanding into a broader and more intricate scope.
%
Currently, these two paradigms are converging, as demonstrated in Figure~\ref{fig:Metaverse}.
DT, originating from the industrialization with a focus on manufacturing production, is progressing towards globalization; in contrast, the metaverse, rooted in games and media with a focus on constructing relationships, is moving towards industrialization.

Our extension of DT into the metaverse for DT-SA in EIoT \textbf{brings several advantages}:
\begin{enumerate}[-]
\item (\textbf{To researchers}) Metaverse emphasizes ``{relationships},'' whereas DT primarily focuses on ``{Entities}.''
\item (\textbf{To dispatchers}) Metaverse utilizes {virtual simulation technology to create a virtual world} supporting ``what if'' discussions, whereas DT pursues an image as close to the real world as possible.
\item (\textbf{To patrol inspectors or system operators}) In the metaverse, terminal users are not only information demanders but also information creators.
\end{enumerate}
\textbf{These superiorities enhance DT's capabilities in handling diverse, complex, and uncertain scenarios in EIoT}.

\subsection{Redefined DT for DT-SA in EIoT}

The redefinition of DT is meaningful to DT-SA in EIoT.
On one hand, it serves as an interface for existing knowledge and experience, thereby narrowing the search space of BM, thus enhancing its efficiency, interpretability, robustness, and generalization performance.
On the other hand, it facilitates systematic deduction across various time-scales, evolution paths, and uncertainties, \textbf{liberating simulations from the constraints of ``reality'' to effectively address ``what if'' dilemma before practical implementation}.

DT integrates mechanistic knowledge and digital technology to model the abstract properties of Physical Entities (PEs), encompassing characteristics, behavior, and configuration, thus forming corresponding Virtual Entities (VEs). These VEs enable ergodic simulation, providing nearly full-coverage data for the systematic exploration of "what if" scenarios and other considerations, especially in the context of DERs aggregation or broader EIoT scenarios.

DT utilizes both mechanistic knowledge and digital technology to model the abstract properties of PEs, such as characteristics, behavior, and configuration, thus forming corresponding VEs.
These VEs  enable ergodic simulation, providing nearly full-coverage data for the systematic exploration of ``what if'' scenarios and other considerations, especially in the context of DERs aggregation or broader EIoT scenarios.

Furthermore, the generated data become the cornerstone for constructing a data-intensive BM under the Fourth Paradigm.
From this sense, our DT actively examines real-time sampling data with the assistance of the built-in BM, inferring indirect indicators (e.g,. high-dimensional statistics, deep features) for some domain-specific task.
These indicators allows us to {evaluate the current situation, learn from past experience, and forecast future trend}.
Overall, the redefined DT is dedicated to \textbf{systematically implementing SA of VEs} throughout their life-cycle in the digital space, considering associated uncertainty, risk, and opportunity.

\subsection{Framework and Roadmap of DT-SA}
\label{Sec:Ingredient}

\begin{figure}[htbp]
\centerline{
\includegraphics[width=.48\textwidth]{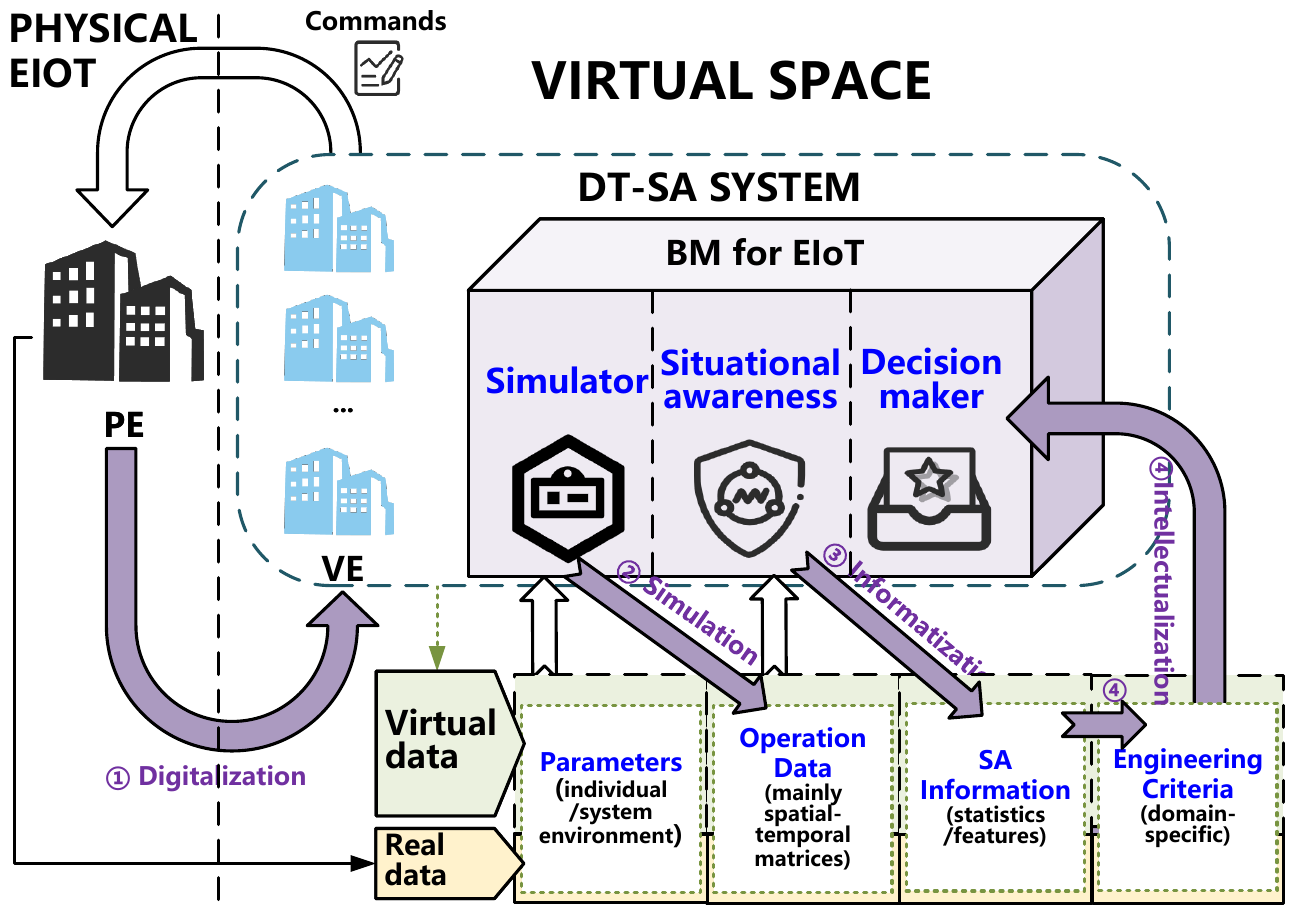}
}
\caption{Schematic of DT-SA framework and its four steps: digitalization-simulation-informatization-intellectualization. }
\label{fig:Foursteps}
\end{figure}

This part, building upon our redefined DT and referencing our prior work~\cite{he2022spatial, he2023situation}, introduces a DT-SA framework, as illustrated in Figure~\ref{fig:Foursteps}.

In general, our DT-SA framework comprises four key steps:
\begin{enumerate}[a)]
  \item \textbf{Digitalization}: the unified characterization of diverse PEs under the condition of mutual operability between PE units in the real world and VE models in the digital space;
  \item \textbf{Simulation}: the generation of a batch of multi-scenario dataset in the digital parallel space through intensive simulation using VE models;
  \item \textbf{Informatization}: the extraction of high-dimensional statistics or deep features from the dataset through BDA or AI for BM training;
  \item \textbf{Intellectualization}: the customization of engineering criteria to assist decision-making for domain-specific tasks.
\end{enumerate}

\begin{figure*}[htbp]
\centerline{
\includegraphics[width=.94\textwidth]{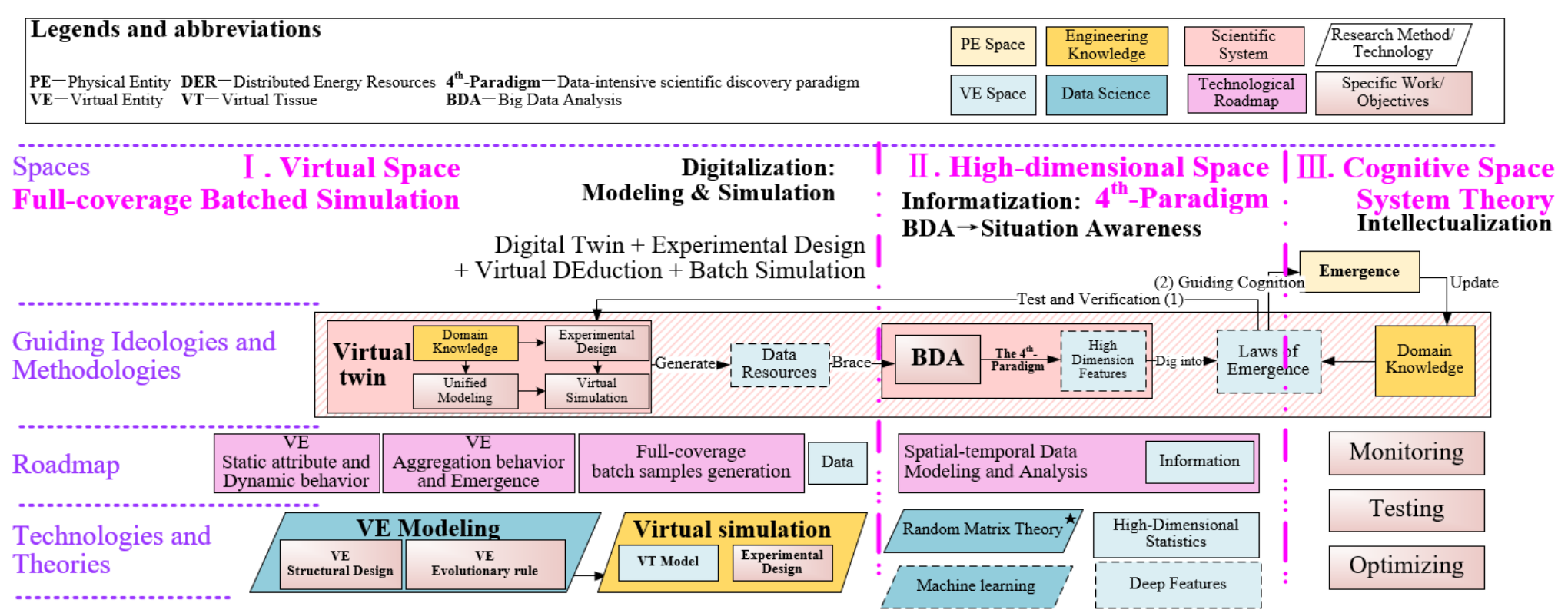}
}
\caption{Ecosystem contributing to DT-SA's evolution and effectiveness}
\label{fig:VirtualLearn}
\end{figure*}

In relation to these four steps, an ecosystem is suggested as depicted in Figure~\ref{fig:VirtualLearn}, contributing to DT-SA's evolution and effectiveness.
This ecosystem explores three interconnected spaces: a) \emph{virtual space} for digitalization and simulation, b) \emph{high-dimensional space} for informatization, and c) \emph{cognitive space} for intellectualization.
\textbf{The interdependence among the three spaces is illustrated through the flow of data}. The output of \emph{virtual space} (batched sampled data) is the input of the \emph{high-dimensional space}, and the output of the \emph{high-dimensional space} (informative high-dimensional statistics and deep features) is the input of the \emph{cognitive space}. Finally, engineering indicators are customized in \emph{cognitive space} for decision-making on domain-specific applications.

\subsubsection{Virtual Space (Space \uppercase\expandafter{\romannumeral 1})}
Manual/history (labelled) data are utilized to build VE models by pre-learning, and to build or update the BM in the (early) stages of DT.
Space~\uppercase\expandafter{\romannumeral 1} takes data resources as input, generating full-coverage sample data and decisions as outcomes.
It involves two major components---{DT} as data engine and {BM} as intelligence engine.

\subsubsection{High-dimensional Space (Space \uppercase\expandafter{\romannumeral 2})}
This space strives to transform the presented system, characterized by low-dimensional raw data, into a perceived system with informative statistics and features.
{Data utilization methodology}, involving data modeling and analysis, are at the heart.

\subsubsection{Cognitive Space (Space \uppercase\expandafter{\romannumeral 3})}
Manual intervention and ultra-realistic actuators are employed for {virtual test} on domain-specific decisions within the metaverse.
It provides a {channel} for evaluating ``{what if}'' scenarios under normal or urgent conditions, in order to pick out or even to generate reliable/prescient decisions.

Additionally, feedback  establishes a {closed-loop} among these three spaces, allowing us to update BM or to correct/adjust VE models in a continuous cycle.
With the closed-loop feedback attribute and successive iteration mechanism, DT enables simulations of {new ideas} that can be tested to evaluate the environmental impact and to confirm the performance in the seamless digital space, {before giving an  implementation in practice}.
Furthermore,  \textbf{continuous optimization of each module/algorithm} within the DT-SA ecosystem can be carried out in this successive iteration process, ensuring the \textbf{nearly real-time efficiency and adaptability} of DT-SA in the complex EIoT.

\section{Modelling and Simulation of Virtual Entity}
\label{Sec:VEModel}

This section, aligned with our designed DT-SA framework in Section~\ref{Sec:Ingredient}, establishes a unified and refined VE model for PE characterization.
This VE model not only mirrors real-world PEs, but also possesses the capability to self-evolve in the metaverse.
Additionally, VE modeling is informed by systems theory, which advocates for a holistic approach to studying the entire system, placing emphasis on system's openness, self-organization, and hierarchical structure.

\subsection{Demanded Attributes of Virtual Entities}

The VE model is formulated to describe PE's abstract properties (e.g., characteristics, behavior, configuration, and performance) in a knowledge-data hybrid manner.
As discussed in Section~\ref{Sec:BM}, three models for VE are considered:  a) white mechanism model; b) black deep network model; and c) statistical model.
Outlined below are key features expected of the VE model:
\normalsize{}

\begin{enumerate}[-]
\item \textbf{Generality}: VE is applicable to most PEs in existence, encompassing various types, services, and levels.
\item \textbf{Representability}: VE can ``statically'' profile PE by reflecting its basic properties, functions, demands, and other attributes.
\item \textbf{Predictability}: VE can ``dynamically'' simulate PE's evolutionary behavior on multiple time scales, deducing consequences with the help of incorporating domain-specific knowledge and data science.
\item \textbf{(Real-time) Interactivity}: VE can exchange data, information, and energy with exteriors (PEs, other VEs, and the system) in nearly real time.
\item \textbf{Interoperability}: Closure topologies and mathematical operators can be defined to study the interaction of multiple environmental VEs.
\item \textbf{Subjectivity}: VE can simulate motivation and decision-making processes to derive the "right" decision.
\item \textbf{Self-evolution}: VE can self-learn, self-improve, and self-update through accumulated data.
\end{enumerate}

\subsection{Designing of Virtual Entity Model}
\label{sec:VEM}
To incorporate the aforementioned features, the VE model is structured as follows:
\[\textbf{VE} = (\textbf{VE}_\text{Attr}, \textbf{VE}_\text{Perf}, \textbf{VE}_\text{Info}, \textbf{VE}_\text{Agg},  \textbf{VE}_\text{SA})\]
where
\begin{enumerate}[-]
  \item $\textbf{VE}_\text{Attr}$: PE's natural/social attributes, including its components, resources and demand, adjustable volume, network structure/parameters, and so on.
  \item $\textbf{VE}_\text{Act}$: PE's optional action, such as energy generation/consumption, which depends on $\textbf{VE}_\text{Info}$ and $\textbf{VE}_\text{SA}$.
  \item $\textbf{VE}_\text{Info}$: Information from the inner and exterior of PE.
  \item $\textbf{VE}_\text{Agg}$: Aggregation/decomposition of PEs, relevant to the entire procedure including its motivation, decision, action, and consequence, with an emphasize on the emergence of intelligence \& value (see Section~\ref{Sec: motivationofAgg}).
  \item $\textbf{VE}_\text{SA}$: DT-SA module, taking  $\textbf{VE}_\text{Attr}$ \& $\textbf{VE}_\text{Info}$ as inputs, and  $\textbf{VE}_\text{Agg}$ \& $\textbf{VE}_\text{Act}$ as outputs.
\end{enumerate}

In such a manner, the VE model is designed to map the PE's operations in the physical world and its responses to the external environment.
For instance, if $\textbf{VE}_\text{Info}$ module anticipates an upcoming peak in power consumption, $\textbf{VE}_\text{Attr}$ module will be activated to assess potential flexible resources, in order to assist $\textbf{VE}_\text{SA}$ module in making demand-response decisions for $\textbf{VE}_\text{Act}$.

Our study on the VE model emphasizes aggregation behavior, with three indicators describing motivation, participants, and effectiveness of aggregation:
\begin{enumerate}[-]
  \item \textbf{Aggression degree}:  tendency towards change or evolution, largely influenced by $\textbf{VE}_\text{Attr}$. A flexible and intelligence unit has a high level of aggression degree.
  \item \textbf{Fitness degree}:  VE's fitness for other VEs or the environment, depending on $\textbf{VE}_\text{Attr}$ and $\textbf{VE}_\text{Info}$. Energy storages, for example,  have a high fitness degree with DGs. For rigorous quantization of this degree, a ``binary operation'' and ``distance'' should be defined.
  \item \textbf{Emergence degree}:  benefit as the consequence of aggregation, such as emergence phenomena of new characteristics, ability, and behavior, or the alleviation of uncertainty caused by the mixing of DGs and energy storages.
\end{enumerate}

\subsection{Quick Start and Self-evolution of DT-SA with VE}
\label{Sec:QuickStart}

Creating an omniscient DT that covers every facet of EIoT is both impractical and unnecessary, especially in the near future.
For domain-specific engineering tasks, a sub-DT---essentially a condensed version of the omniscient DT---is often adequate.
For example, when the EIoT is on the verge of instability, economic considerations might take a backseat.

In practice, engineers may  initiate a quick DT by creating simple VEs, offering early-cycle technical support in the program.
Once deployed, these VEs, simple at first, may evolve over time, driven by the accumulation of operating data or through the feedback from manual interventions.
This quick start capability enhances DT-SA's adaptability to the complex EIoT.
In general, instead of adhering to a fixed model, the VE embraces a self-adaptive, self-learning, and self-evolve model with an iterative process: modeling$\rightarrow$deployment$\rightarrow$testing$\rightarrow$updating$\rightarrow$remodeling$\cdots$.

\section{Case Studies of DT-SA in EIoT}
\label{Sec:case}

This section introduces practical applications of DT-SA through three case studies. These cases demonstrate how the framework is applied in real-world engineering projects and the development of relevant standards, and verify  the superiority and necessity of our proposed DT-SA.

\subsection{Case\#1 DT-Based Substation Equipment Fault Diagnosis}
\label{sec:Case1}
\begin{figure}[!ht]
  \centering
  \includegraphics[width=0.485\textwidth]{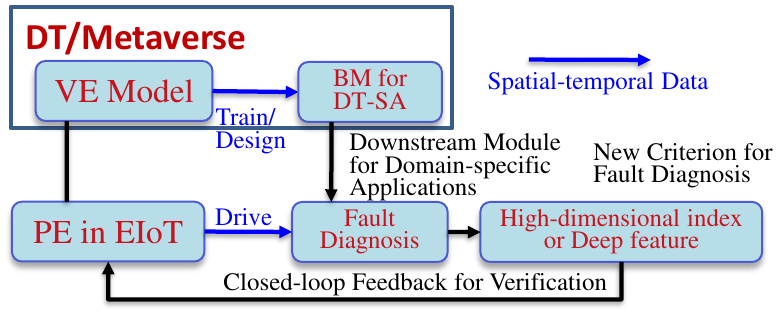}
  \caption{DT-SA for Fault Diagnosis}
  \label{fig:DTFault}
\end{figure}

This part, as depicted in Figure~\ref{fig:DTFault}, by taking fault diagnosis as a domain-specific SA task, makes a comparison among traditional mode, DT mode, and metaverse mode.
The result demonstrates how far our efforts have come.

\subsubsection{Engineering Background and Classical Indicator}
{\text{\\}}

\begin{figure}[!ht]
  \centering
  \includegraphics[width=0.49\textwidth]{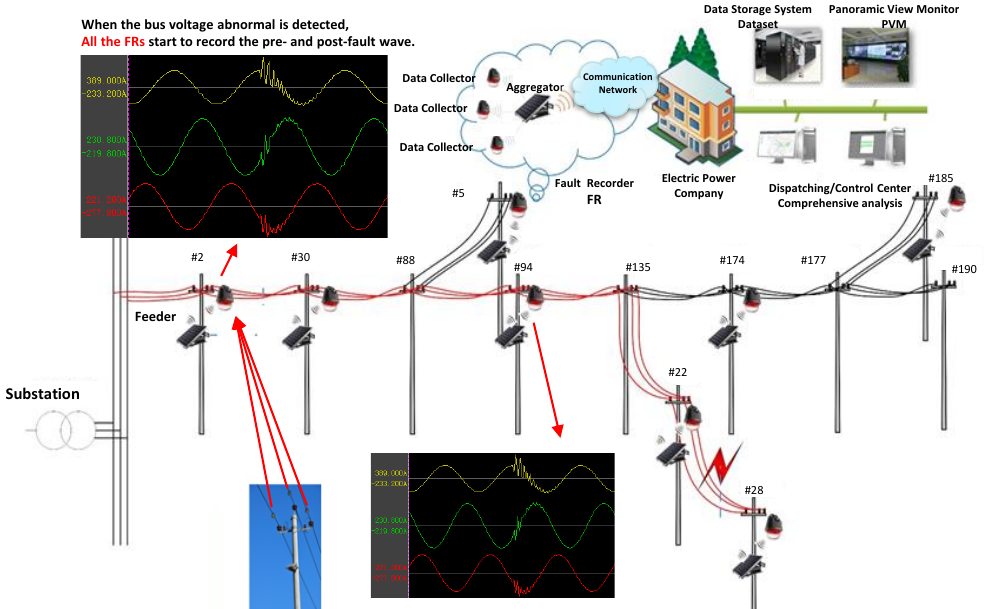}
  \caption{Waveform of $V$ and $I$ in a real distribution network}
  \label{fig:Distri}
\end{figure}

Small current ground fault, often with insignificant  fault characteristics, is a common fault in the distribution network.
Several measurements, such as fault recorder (FR), $\mu$PMU, and PT/CT (potential transformer/current transformer), are deployed for sampling $V$ and $I$  on heterogeneous buses, as illustrated in Fig.~\ref{fig:Distri}.

Classical indicator for this type of fault is zero-sequence current, which is a low-dimensional indicator depended on only 3-dimensional raw data, i.e., $I_\text{A}, I_\text{B}, I_\text{C}$ confined to each single FR, as depicted in center part of Figure~\ref{fig:DTVSModel}.
There are several disadvantages for  zero-sequence current indicator: a) FR is vulnerable to external environment, communication and other factors, its own accuracy and precision is not highly reliable; b) Considering both sensitivity and reliability, the fault threshold is difficult to set---a high threshold is unreliable and prone to missed alarms; conversely, it is too sensitive and prone to trigger false alarms.
In general, zero-sequence current indicator does not make full use of the spatial-temporal data (from heterogeneous lines) and their correlation..

\noindent \subsubsection{DT-SA Indicators for Fault Diagnosis}
{\text{\\}}

\begin{figure}[htb]
\centering
\subfloat[Raw Data \& LES $\tau_{1-42}$]{\label{fig:cs1}
\includegraphics[width=0.48\textwidth]{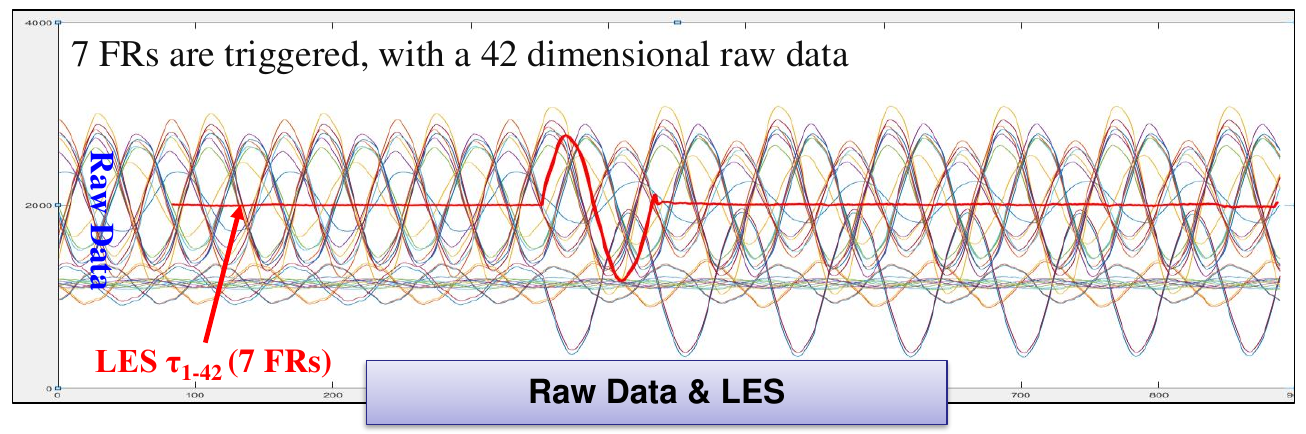}}

\subfloat[LES $\tau_{1-42}$ V.S. Zero-sequence Current (Sensitive)]{\label{fig:cs2}
\includegraphics[width=0.48\textwidth]{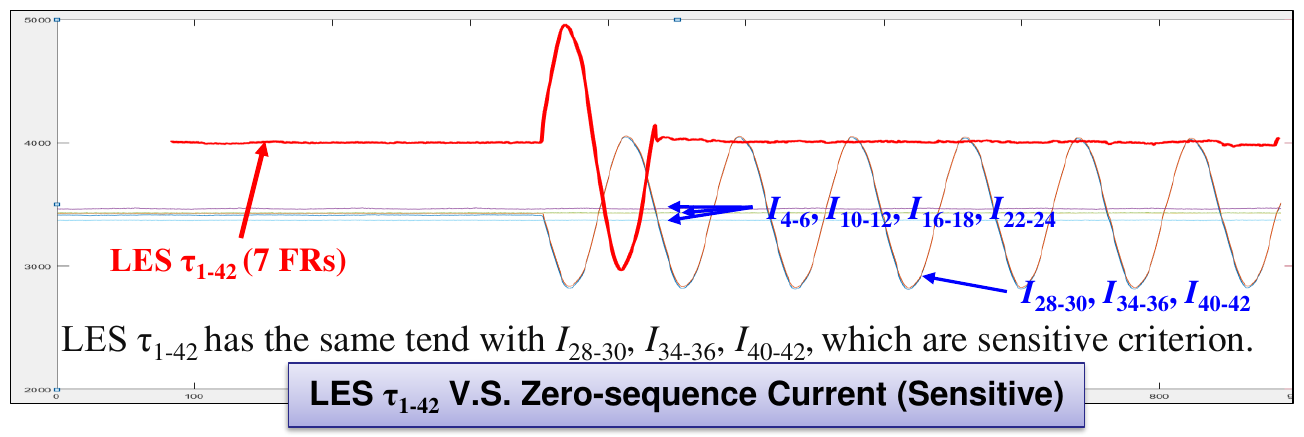}}

\subfloat[LES $\tau_{1-24}$ V.S. Zero-sequence Current (Insensitive)]{\label{fig:cs3}
\includegraphics[width=0.48\textwidth]{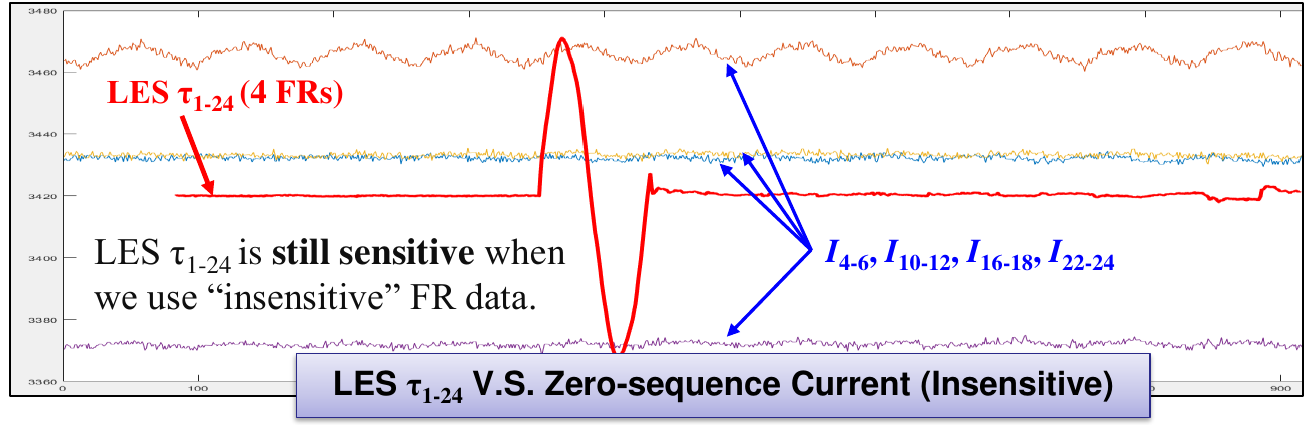}}

\caption{Test and improve the engineering indicators in virtual digital world: classical indicators V.S. LES indicator}
\label{fig:Case}
\end{figure}

We take a real case using field data in some distribution network.
When a fault occurs, 7 FRs are turned on to record 42-dimensional data (each FR records 6-dimensional data: three-phase voltage, i.d., $1,2,3$, and three-phase current, i.d., $4,5,6$).
The FR records 10 cycles before and after its start-up, covering the life-cycle of the fault, as shown in Figure~\ref{fig:cs1}.

On one hand, a built-in module for each FR, separately, calculates its corresponding classical zero-sequence current indicator (7 indicators in all). Without loss of generality, suppose they are $I_{4-6}, I_{10-12}, I_{16-18}, I_{22-24}, I_{28-30}, I_{34-36}, I_{40-42}$, and
the classical indicators are sensitive to the fault for the last 3 FRs ($I_{28-30}, I_{34-36}, I_{40-42}$), and are insensitive for the first 4 FRs ($I_{4-6}, I_{10-12}, I_{16-18}, I_{22-24}$).

On the other hand, the 42-dimensional spatial-temporal data, in the form of $N \times T$ matrix, compose the VE model of the life-time of the fault (pre-fault, post-fault).
We get to look at the data from multiple FRs as a matrix, and implement spatial-temporal analysis to the matrix, as depicted in the right part of Figure \ref{fig:DTVSModel}.
In concrete, we, according to Eq.~\ref{eq:DDLES}, extract a high-dimensional spectral indicator LES based on RMT as a novel indicator, i.e., $\tau_{1-42}$.
The proposed LES $\tau_{1-42}$ indicator does make use of the spatial-temporal correlation of all the 7 strongly associated FRs. Figure~\ref{fig:cs2} shows that the LES indicator $\tau_{1-42}$ has the same trend as classical indicators $I_{28-30}, I_{34-36}, I_{40-42}$, which are also sensitive indicators.

Then we move to the digital space (metaverse). If we suppose that the data from 3 sensitive FRs are unreliable or even totally missing, as shown in Figure~\ref{fig:cs3}, then we can not make a correct SA about our distributed network using the left classical zero-sequence current indicators ($I_{4-6}, I_{10-12}, I_{16-18}, I_{22-24}$).
However, the LES indicator  $\tau_{1-24}$, which can be calculated out using 24-dimensional data only from the left 4 (insensitive) FRs, is still competent in the SA task.
In such a manner, our DT-SA offers a data-intensive discovery for fault diagnosis application.
This fault diagnosis module has already been deployed to real systems, and it exhibits better performance.

In addition, this case tells how to test our VE in metaverse.
Our case says that our LES indicator  $\tau_{1-24}$ is still effective when all the 3 sensitive FRs are unreliable.
Then it is reasonable to believe that our algorithm is still valid for several arbitrary FR data missing/anomalous, and hence we need not test these scenarios in our metaverse.

\subsection{Case\#2 DT-SA Project and Relevant Standards for EIoT in Lingang, China}

The implementation of EIoT stands as a strategic initiative in alignment with sustainable development and a low-carbon economy.
In Lingang New District, Shanghai, China, a prominent project exemplifies the implementation of our DT-SA framework within the international free-trade zone.

\subsubsection{Background and Overview of Lingang DT-SA Project}
{\text{\\}}

Lingang is endowed with abundant resources, making it an ideal locale for investigating the behavior of  DERs within a complex EIoT.
Lingang has a rich array of climate-susceptible natural energy resources, along with diverse users with varying resources, characteristics, and demands.
\begin{enumerate}[-]
    \item \textbf{Natural Energy Resources}---approximately 500MW (installed capacity) wind turbine, 200MW photovoltaics, etc.
    \item \textbf{Demand Response Users}---electric vehicle (EV) charging stations with a considerable scale, etc.
    \item \textbf{Important users}---integrated circuit manufacturers, electric vehicle manufacturers, and large aircraft manufacturers, etc.
\end{enumerate}
The deliberate and interconnected deployment of these elements positions Lingang as a compelling exemplar.

\begin{figure*}[!ht]
  \centering
  \includegraphics[width=0.88\textwidth]{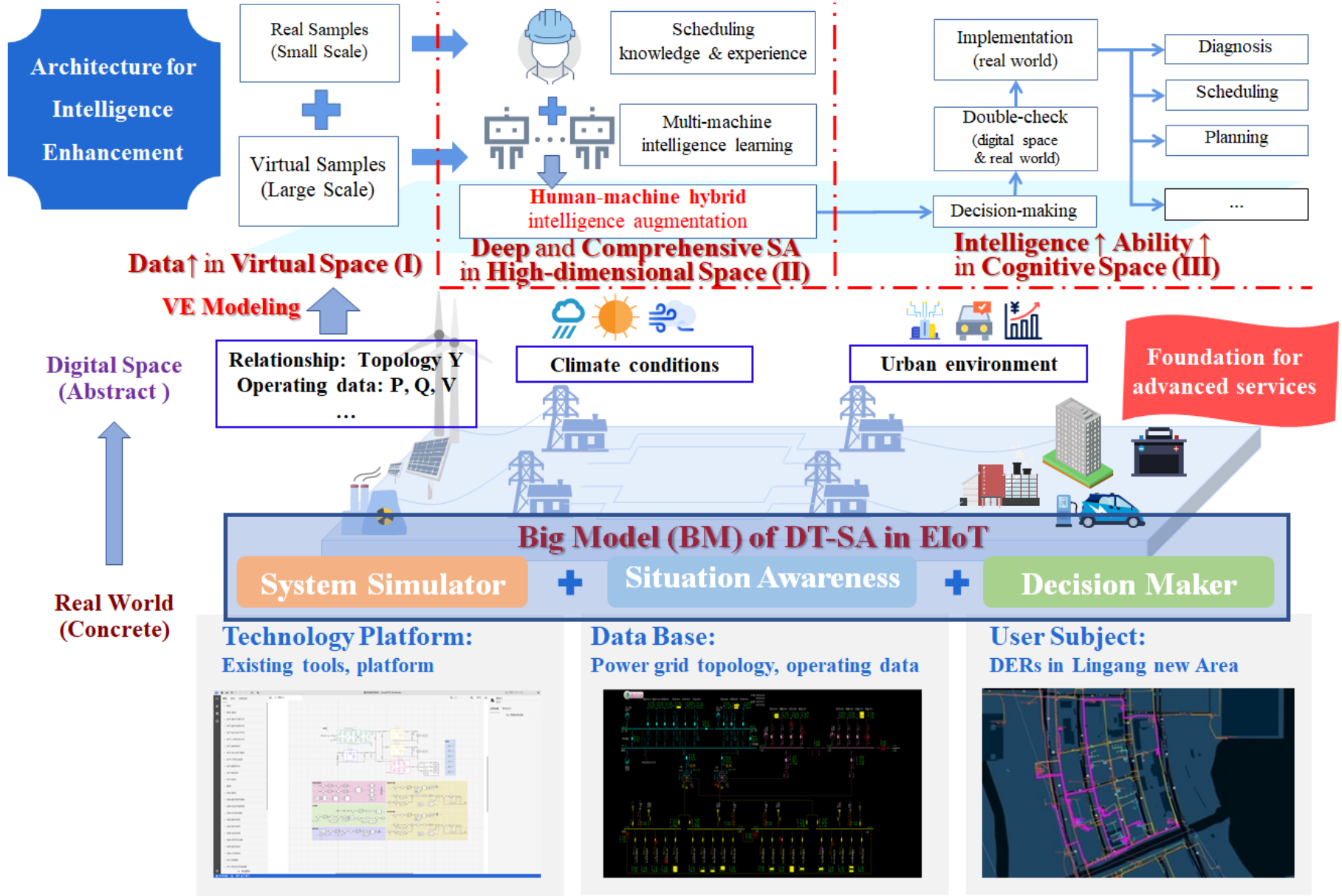}
  \caption{Demonstration scheme of DT-SA Framework for EIoT in Lingang, China}
  \label{fig:EIoTLG}
\end{figure*}

In this context, the collaborative efforts between State Grid Shanghai Municipal Electric Power Company (Author 2) and Shanghai Jiao Tong University (Authors 1, 3, 4) initiated the Lingang DT-SA Project for EIoT within the Lingang regional distribution network.
This project is funded by the Science and Technology Commission of Shanghai Municipality (Project No: 21DZ1208300, 2021.11$\sim$2024.11), and titled ``Research and Application of Situation Awareness and Efficient Operation Technology for EIoT in a Digital City''.
Its overarching goal is to contribute to the advancement of city digitalization, energy optimization, and urban management.
Figure~\ref{fig:EIoTLG} illustrates this project.

The project develops a digital infrastructure for managing DERs within EIoT, leveraging our proposed DT-SA framework (refer to Figure~\ref{fig:Foursteps}).
The DT-SA Framework seamlessly integrates technology platform, data base, and user subjects to organically construct the EIoT-DT and a digital space for the Lingang metaverse.
It is noteworthy that the construction of the Lingang metaverse aligns with the motivations and engineering principles of the DT-SA framework, as illustrated in Figure~\ref{fig:Redu}---for instance, we may conceptualize the EV user as an EM (energy molecule), the EV charging stations as an EC (energy cell), and the virtual power plant (VPP~\cite{yin2022trading}) containing these EV charging stations as an ET (energy tissue).
The behaviors of the above DERs and their aggregation in the metaverse (corresponding VM, VC, VT) are influenced by factors such as traffic condition and climate environment.
Utilizing the Lingang metaverse, we can conduct a comprehensive and systematic exploration on DERs behaviors and their impact on the EIoT in Lingang with minimal trial-and-error costs.

\subsubsection{DT-Based Automatic Control of Air Conditioning}
{\text{\\}}

\begin{figure*}[htb]
\centering
\subfloat[DT-based automatic control of air conditioning]{\label{fig:automatic}
\includegraphics[width=0.92\textwidth]{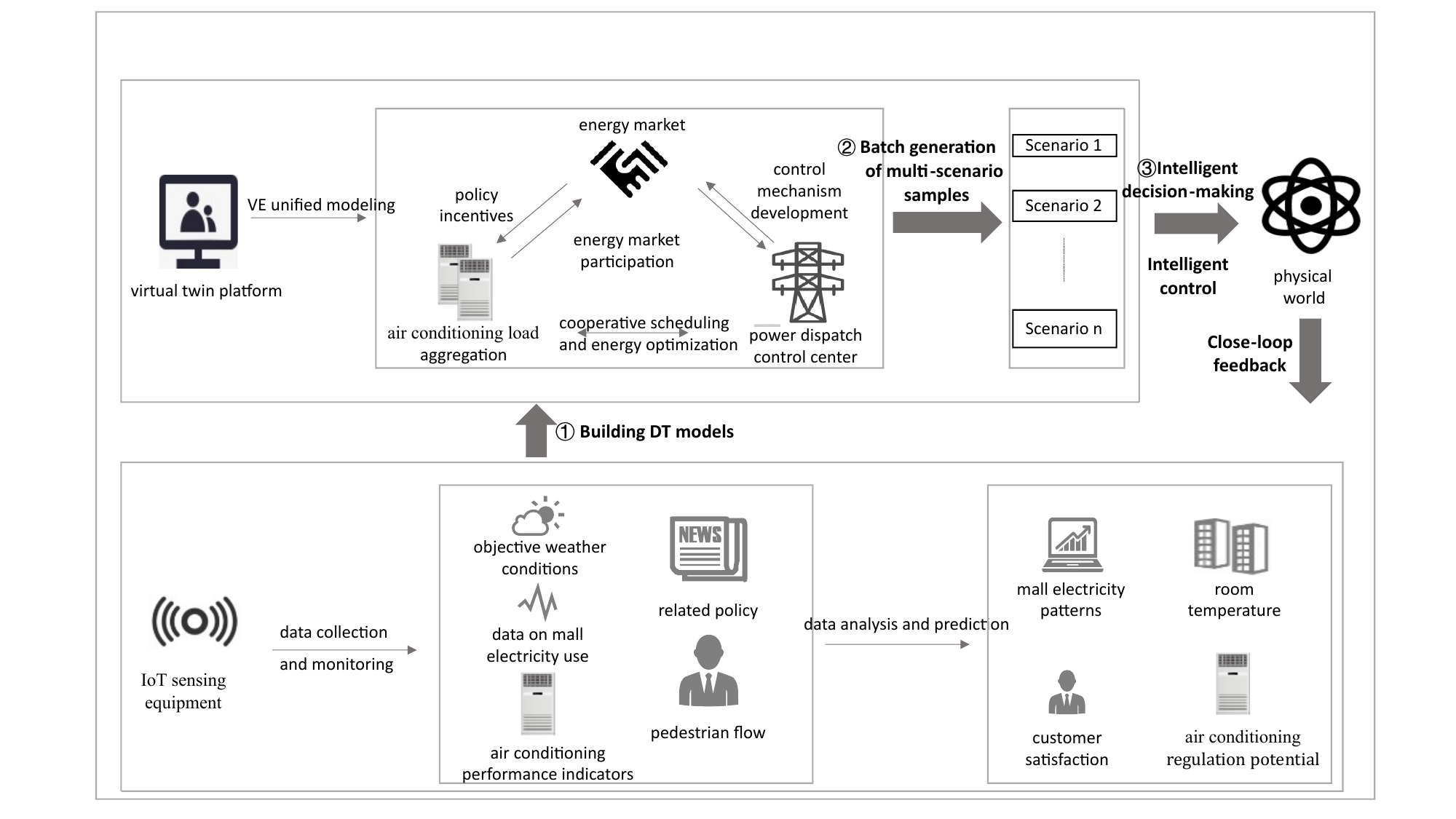}}

\subfloat[Use Case: electrical vehicle]{\label{fig:case22}
\includegraphics[width=0.40\textwidth]{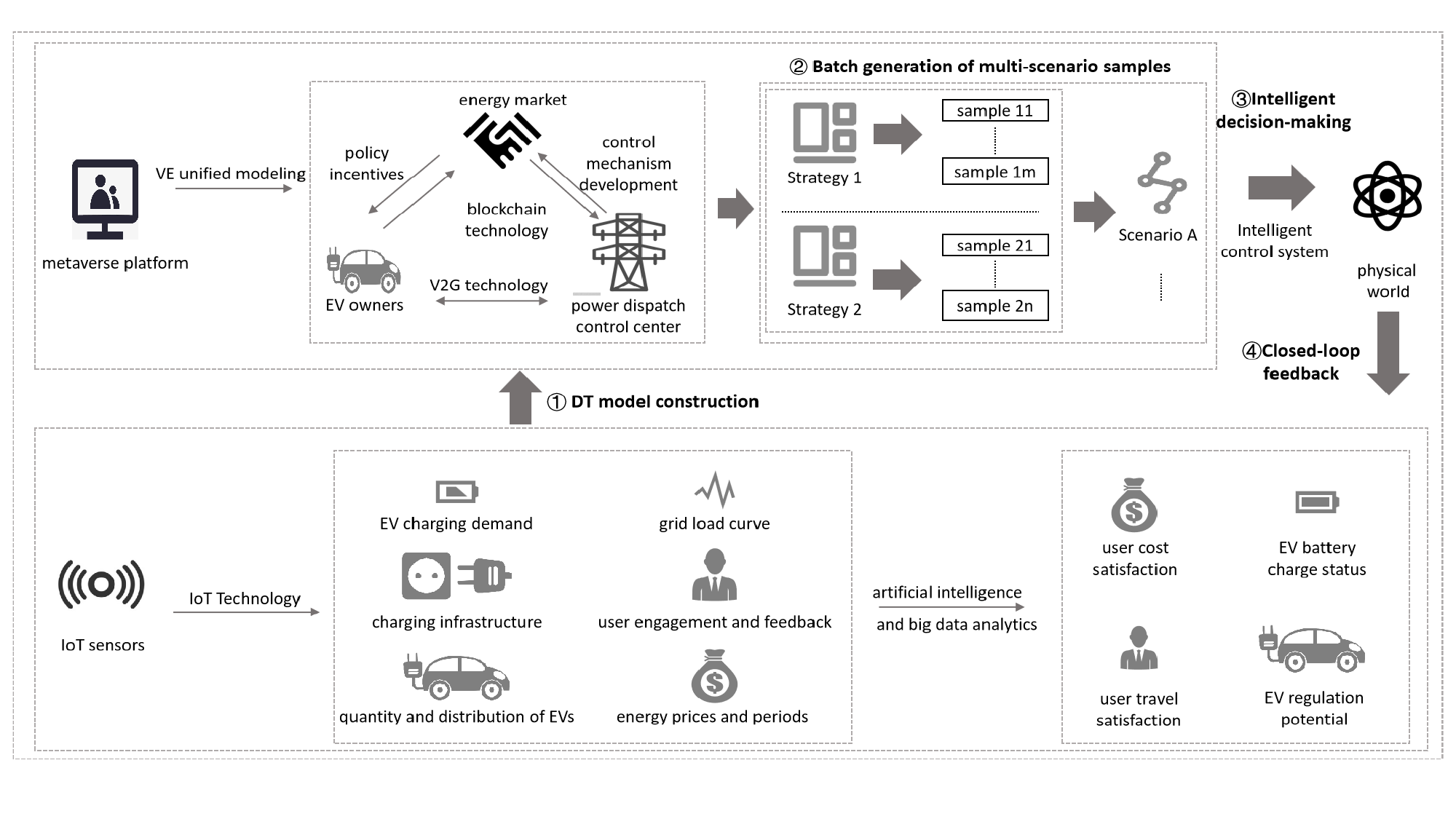}}
\subfloat[Use Case: energy storage]{\label{fig:case33}
\includegraphics[width=0.40\textwidth]{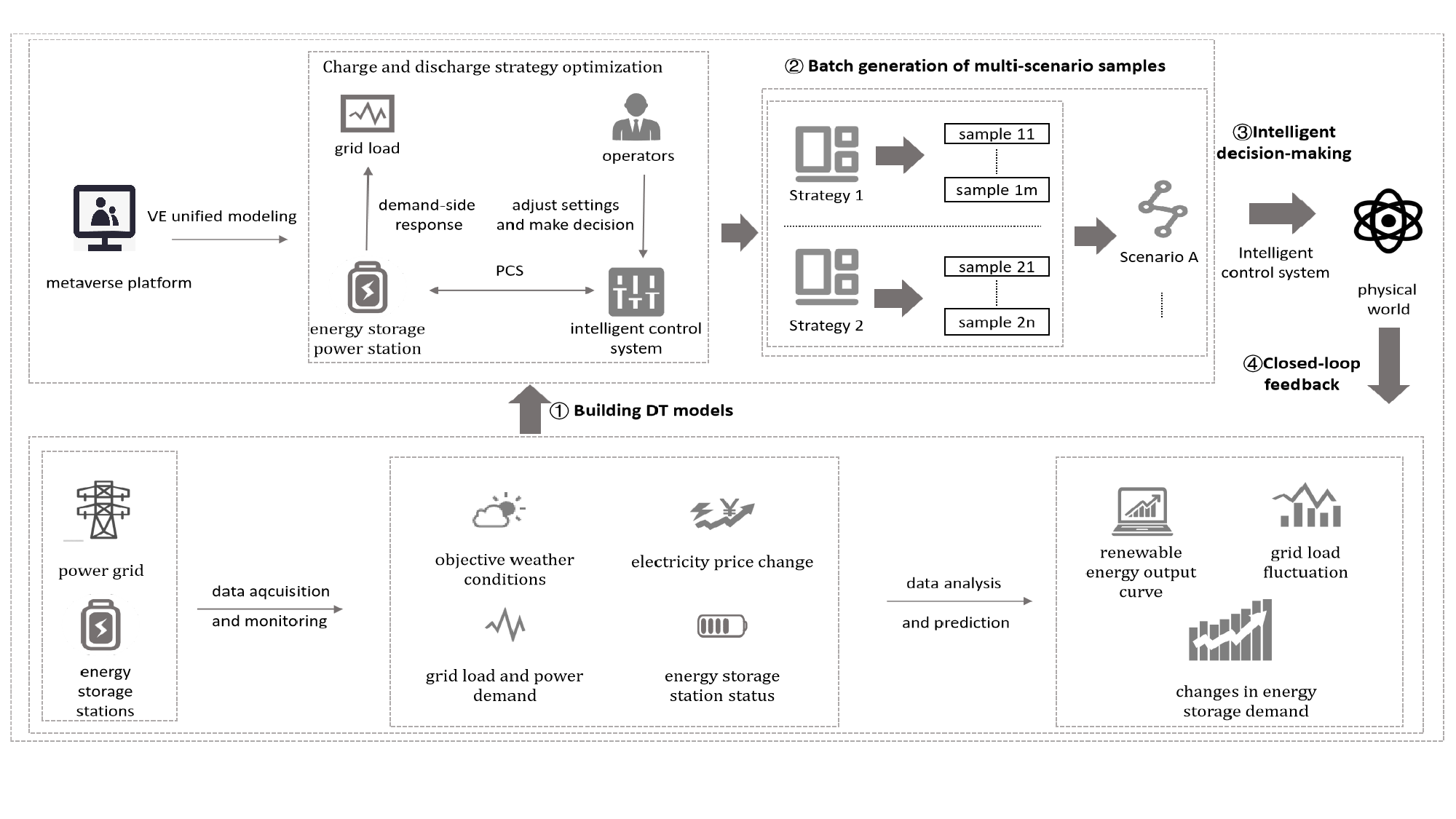}}

\caption{Power Metaverse: Use Cases}
\label{fig:CaseEEE}
\end{figure*}

Illustrating the effectiveness of the DT-SA framework, we focus on the automatic control of air conditioning (refer to Fig.~\ref{fig:automatic}) within the broader context of the Lingang Project.
Air conditioning load, recognized as a large-scale adaptable demand-side resource, has the capacity for conducting load response (as either an independent entity or part of a VPP), contributing to EIoT dispatching strategies such as load peak shaving or network congestion management.

Our project establishes a sub-DT dedicated to air conditioning by mapping real-world parameters (climate, pedestrian flow, set temperature, etc.) into the digital space.
Within this DT system, various air conditioning output strategies are simulated, including load shifting (Strategy 1), load reduction (Strategy 2), etc.
Utilizing those batch-generated sample data, the DT-SA platform, with the help of BM, determines the optimal strategy.
In such a manner, we give a comprehensive consideration on the uncertainty of environment and interests of multiple-stakeholder, balancing the needs of power dispatchers (load peak shaving), mall customers (thermal comfort experience), and mall operators (minimizing electricity expenses).
This case reflects the superiority of our DT-SA framework in DERs management in EIoT.

\subsubsection{Development of Relevant Standard---Power Metaverse}
{\text{\\}}

The Lingang DT-SA Project has introduced an ITU technical standard titled ``Power Metaverse: Use Cases Relevant to Grid Side and User Side'' (refer to~\cite{PowerMeta2023HE}).
This report provides a comprehensive overview of the power metaverse, presenting a roadmap and detailing seven use cases, including the air conditioning case discussed earlier.
These cases showcase the application of the power metaverse in the power system, addressing both user and grid perspectives.
Each use case offers a thorough description of the application scenario, including assumptions, service scenarios, and requirements.
\begin{enumerate}[-]
    \item The user-side cases illustrate how the metaverse can be employed to balance the supply and demand of the power grid by promoting the participation of adjustable and controllable resources on the user-side, including air conditioning, electric vehicles, combined heat and power, energy storage, and multienergy complementarity; these cases resemble scenarios like air conditioning control, as depicted in Figures~\ref{fig:case22} and \ref{fig:case33}.
    \item The grid-side cases illustrate how the metaverse can be employed to enhance the automation and intelligence of core businesses in the distribution network, including resilient power grid, and DGA (dissolved gas analysis) inspection.
\end{enumerate}

The specification outlined in this technical report serves as a valuable reference for intelligent decision-making, technical research, and practical applications, underscoring the significance of the Lingang DT-SA framework in both engineering applications and theoretical guidance.

\section{Conclusion}
\label{Sec: Conclu}

This work introduces a cutting-edge redefinition of DT in the context of Fourth Paradigm, CAS, and BM.
The redefined DT extends beyond its original form, transforming from a limited  ``1-1 mapping'' mirror world towards an expansive ``free'' metaverse.
The redefined DT, evolving from a confined ``1-1 mapping'' mirror world towards a expansive ``free'' metaverse, enables us to proactively address hypothetical scenarios before practical implementation, thereby reducing trial-and-error costs.
Additionally, the metaverse contributes substantially to the BM training by providing extensive virtual data.

Next, we delve into a concise yet comprehensive DT-SA framework, focusing on its role in gaining systematic and quantitative insights into complex systems through spatial-temporal data.
Notably, we emphasize the merits of DT-SA, including its model-free nature, theoretical support from RMT, compatibility with existing knowledge, and resilience against uncertainties and data errors.
This positions DT-SA as a compelling alternative for DT-SA in the complex EIoT.
Positioned as the future development direction of EIoT, DT-SA offers valuable insights and intelligence across various technical aspects (e.g., aggregation and system modeling, analysis of chaotic systems with emergence phenomena), with a coverage of wide applied domains such as fault diagnosis, peak shaving dispatching.

In conclusion, the development of the DT-SA encompasses various aspects spanning industry, academia, and research. Our DT-SA framework is compatible, so it serves as a novel foundation for collaboration among experts from diverse fields. Besides, the DT-SA framework demonstrates adaptability beyond the EIoT realm, showing promise in implementation within other systems, such as intelligent cities and brilliant factories. This broad exploration holds potential benefits for both the engineering community and data science, underscoring the versatility and significance of DT-SA in advancing our understanding and management of complex systems.

\bibliographystyle{IEEEtran}
\bibliography{helx}

\normalsize{}
\end{document}